\def\apjl{ApJ~}%
\def\apjs{ApJS}%
\def\ao{Appl.~Opt.}%
\def\aap{A\&A~}%
\documentclass[preprint,12pt]{elsarticle}

\usepackage[normalem]{ulem}

\usepackage[left=2cm,right=	2cm,top=2cm,bottom=2cm]{geometry}
\usepackage{subcaption}
\usepackage{float}
\usepackage{graphicx}
\usepackage{xcolor}
\usepackage{hyperref}
\journal{Icarus}

\begin{document}

\begin{frontmatter}



\title{Optical observations and dust modelling of comet 156P/Russell-LINEAR}

\author[inst1,inst2]{K.Aravind}

\affiliation[inst1]{organization={Physical Research Laboratory},
            addressline={Navarangpura}, 
            city={Ahmedabad},
            postcode={380058}, 
            state={Gujarat},
            country={India}
}

\author[inst1,inst3]{Prithish Halder}
\author[inst1]{Shashikiran Ganesh}
\author[inst3]{Devendra Sahu}
\author[inst4,inst5]{Miquel Serra-Ricart}
\author[inst6]{José J. Chambó}
\author[inst3]{Dorje Angchuk}
\author[inst3]{Thirupathi Sivarani}

\affiliation[inst2]{organization={Institute of Technology Gandhinagar},
            addressline={Palaj}, 
            city={Gandhinagar},
            postcode={382355}, 
            state={Gujarat},
            country={India}}
         
\affiliation[inst3]{organization={Indian Institute of Astrophysics},
            addressline={Koramangala}, 
            city={Bangalore},
            postcode={560034}, 
            state={Karnataka},
            country={India}}

\affiliation[inst4]{organization={Instituto de Astrof$\acute{i}$sica de Canarias},
            addressline={C/Vía Láctea s/n}, 
            city={La Laguna},
            postcode={E-38205}, 
            state={Canarias},
            country={Spain}}

\affiliation[inst5]{organization={Departamento de Astrof$\acute{i}$sica},
            addressline={Universidad de La Laguna}, 
            city={La Laguna},
            postcode={E-38205}, 
            state={Canarias},
            country={Spain}}
\affiliation[inst6]{organization={Cometografia.es},
            city={Vall$\acute{e}$s}, 
            postcode={E-46818}, 
            state={Valencia},
            country={Spain}}

\begin{abstract}
Comet 156P/Russell-LINEAR is a short period Jupiter family comet with an orbital period of 6.44 years. The results from spectroscopic, photometric, polarimetric observations and dust modelling studies are presented here. From the spectroscopic study, strong emissions from $CN (\Delta \nu = 0)$, $C_3 (\lambda4050$\AA), $C_2 (\Delta \nu = +1)$ and $C_2 (\Delta \nu = 0)$ can be observed during both the epochs of our observations. The Q($C_2$)/Q(CN) ratio classifies the comet as a typical comet. The imaging data reveals the presence of jets. The dust emission from the comet is observed to have a non-steady state outflow due to the presence of these strong jets which subside in later epochs, resulting in a steady state outflow. Polarimetric study at two different phase angles reveals the degree of polarization to be comparable to Jupiter family comets at similar phase angles. Localized variations in polarization values are observed in the coma. The dust modelling studies suggest the presence of high amount of silicate/low absorbing material and indicate the coma to be dominated by higher amount of large size grains with low porosity having power law size distribution index = 2.4. The observed activity and dust properties points to a similarity to another Jupiter family comet, 67P/Churyumov–Gerasimenko.  
\end{abstract}

\begin{highlights}
\item We carried out optical spectroscopic, photometric, polarimetric observations and dust modelling of a rarely observed Jupiter-Family-Comet (JFC):  156P/Russell-LINEAR
\item The spectroscopic composition is similar to the typical class of comets and the production rates had increased around perihelion.
\item The activity and dust composition was observed to be similar to that reported for  comet 67P/Churyumov–Gerasimenko.
\item The coma was observed to have a non-steady state outflow of dust emission. Presence of strong jets were also observed. The strength of these dust jets are seen to subside a few weeks after perihelion possibly resulting in a steady-state outflow of dust emission inside the coma.
\item Spatial variation of polarization in the coma implies the possibility of dust jets being a prominent reason to produce localized variation in the density and size of the dust particles.
\end{highlights}


\end{frontmatter}

% \linenumbers

\section{Introduction}
Comets are the primordial remnants of our Solar system, containing pristine materials that were present in the protosolar nebula. Enormous mixing up of these planetesimals occurred during the formation of the giant planets in the Solar system \citep{nice,grandtack}.  Such processes led to the formation of various reservoirs containing the cometary bodies \citep{reservoir_formation}. Intensive studies of essentially all the observable comets belonging to various reservoirs are required to analyse and classify the kind of material present in them. This can help to understand the kind of mixing that may have happened during the formation of the Solar system. Considering the idea that the Jupiter family comets (JFC) have evolved from the Kuiper belt objects \citep{JFC} due to planetary interactions, it is important to study these objects in detail to get a better understanding on the effects of evolution in the material present in the icy bodies \citep{JFC_evolution, farnham_JFCmorphology}.  Simultaneous study of the gaseous and dust emission from the comet, along with polarization analysis can help one gain greater insight into the type of material present in the comet. Similar works have been carried out on a few other comets like C/2014 A4 (SONEAR) \citep{ivanova_C2014A4}, C/2009 P1 (Garradd) \citep{ivanova_C2009P1} and  C/2011 KP36 (Spacewatch) \citep{ivanova_2021}. Comparison between such intensive studies of long period and short period comets would help in unravelling the mystery of primordial and evolutionary signatures present in the comets. \\
In this work, we present the results from spectroscopic, photometric and polarimetric observations along with some dust modelling studies of short period Comet 156P/Russell-LINEAR (hereafter 156P). Comet 156P is a short period Jupiter family comet discovered by Kenneth S. Russell on 1986 September 3. The comet has an orbital period of 6.44 years, a perihelion distance of 1.30 AU and an aphelion distance of 5.6 AU. Comet 156P had a new passage in 2020, with a perihelion on November 17 and the closest approach to Earth (0.38 AU) on 2020 October 29. Despite the multiple passages of the comet through the inner Solar system, there are no reports on  its properties or activity in the earlier literature. The NASA JPL Orbit viewer\footnote{\url{https://ssd.jpl.nasa.gov/tools/orbit_viewer.html}} shows that the comet is well placed for observations during alternate apparitions.  This could be the primary reason for the comet not being properly observed and documented.   The next favourable apparition would be in 2033 with the closest approach to Earth being 0.514 AU. During the 2020 apparition, \citet{156P_Atel_Oct} and \citet{156P_Atel_Nov} reported the activity of 156P in 2020 October and November.  Here we report the pre and post-perihelion spectroscopic activity of the comet 156P, during 2020 October and December, observed from the 2 m Himalayan Chandra Telescope (HCT) of the Indian Astronomical Observatory (IAO) and 1.2 m telescope of the Mount Abu Infrared Observatory (MIRO) respectively. We also combine the imaging data obtained from both HCT as well as the 0.82 m IAC80, Teide observatory telescope to study the evolution of dust jets present in the coma and the variation of the observed Af$\rho$ profile over a period of time. We also report the optical polarimetric observations of the comet observed from MIRO, at two different phase angles. Using the results of the polarimetric observation, dust modelling has been performed with the help of \textsc{Vikram-100 HPC} supercomputing facility at PRL. We briefly describe the observations and data reduction in Section \ref{obs&red}, data analysis and dust modelling in Section \ref{analysis} and hence discuss the main results in Section \ref{results}.

\section{Observations and Reduction}\label{obs&red}
The comet 156P was observed from two Indian observatories namely, Indian Astronomical Observatory (IAO), Hanle, Ladakh and Mount Abu Infra Red Observatory (MIRO), Mount Abu, Rajasthan as well as the Teide Observatory, Tenerife, Spain during multiple epochs.  The 2m Himalayan Chandra Telescope (HCT), at the IAO, was used for imaging and spectroscopy.  At MIRO, the 1.2m telescope was used for spectroscopy and polarimetry. The 0.82 m IAC-80 telescope at the Teide observatory was used for imaging the comet over the period of 2020 November and December. The following sub-sections describe in brief, the details of the observations and data reduction. The observational log, detailing the telescope facility used, observational technique employed, exposure time, number of frames acquired, heliocentric distance, geocentric distance and phase angle  at the time of observations are as given in table \ref{observations}. The NASA JPL HORIZONS\footnote{\url{https://ssd.jpl.nasa.gov/horizons.cgi}} service was used to generate the ephemerides for the comet at all the observing locations.

\begin{table}[h]
\centering
\setlength\tabcolsep{3pt}
\begin{scriptsize}
\renewcommand{\arraystretch}{1.6}
\caption{{Observational Log}}
\begin{tabular}{|c|c|c|c|c|c|c|c|c|c|c|}
\hline
 & & \multicolumn{1}{|c|}{{Telescope}} &  
\multicolumn{1}{|c|}{{Observational}} & Filter/ & & &
\multicolumn{1}{|c|}{{Heliocentric}}   & \multicolumn{1}{|c|}{{Geocentric}} & \multicolumn{1}{|c|}{{Distance scale}} & \multicolumn{1}{|c|}{{Phase}}  \\
\multicolumn{1}{|c|}{{Date}} &  \multicolumn{1}{|c|}{{Time}} & \multicolumn{1}{|c|}{{Facility}}
& \multicolumn{1}{|c|}{{Technique}} & Wavelength & Exposure & \textit{N}
&\multicolumn{1}{|c|}{{Distance} (r$_{H}$)} & \multicolumn{1}{|c|}{{Distance} ($\Delta$)} & \multicolumn{1}{|c|}{{at photo-centre}} & \multicolumn{1}{|c|}{{angle}}    \\
\multicolumn{1}{|c|}{(UT)}  & \multicolumn{1}{|c|}{(UT)} &  & & range ($\AA)$ & (seconds) &  & \multicolumn{1}{|c|}{(AU)}  & \multicolumn{1}{|c|}{(AU)} &    \multicolumn{1}{|c|}{(km pixel$^{-1}$)} &  \multicolumn{1}{|c|}{($^\circ$)}    \\ \hline
 18/10/2020 & 15.21 &HCT$^a$ & Imaging & R & 30 & 5 & 1.380 & 0.483 & 103 & 30.93\\
 18/10/2020 & 15.28 &HCT$^a$& Spectroscopy & 3800-6840 & 1200& 1 & 1.380 & 0.483 & 103 & 30.93\\
 13/11/2020 & 18.70 &MIRO$^b$& Polarimetry & \textit{i} & 10 & 48$^*$ & 1.334 & 0.509 & 266 & 38.90 \\
 18/11/2020 & 21.53 &IAC80$^c$& Imaging & Open & 25 & 32 & 1.333 & 0.526 & 128 & 39.86\\
 21/11/2020 & 00.01 &IAC80$^c$& Imaging & Open & 25 & 32 & 1.334 & 0.537 & 130 & 40.42 \\
 07/12/2020 & 19.08 &IAC80$^c$& Imaging & Open & 25 & 24 & 1.351 & 0.618 & 150 & 42.17 \\
 08/12/2020 & 19.10 &IAC80$^c$& Imaging & Open & 25 & 24 & 1.353 & 0.624 & 152 & 42.25 \\
 14/12/2020 & 19.49 &IAC80$^c$& Imaging & Open & 30 & 28 & 1.367 & 0.663 & 161 & 42.36 \\
 15/12/2020 & 17.28 &MIRO$^d$& Spectroscopy & 3800-7000 & 900 & 1 &1.370 & 0.670 & 158 & 42.35 \\
 16/12/2020 & 14.60 &MIRO$^b$& Polarimetry & \textit{i} & 10 & 48$^*$ & 1.372 & 0.684 & 355 & 42.34 \\
 18/12/2020 & 19.55 &IAC80$^c$& Imaging & Open & 30 & 24 & 1.378 & 0.692 & 168 & 42.32 \\
 20/12/2020 & 23.63 &IAC80$^c$& Imaging & Open & 30 & 20 & 1.384 & 0.707 & 172 & 42.27 \\
 29/12/2020 & 19.36 &IAC80$^c$& Imaging & Open & 25 & 19 & 1.415 & 0.779 & 190 & 41.82 \\
\hline
\multicolumn{11}{|l|}{{$^*$Explained in Section \ref{pol_obs}}}\\
\multicolumn{11}{|l|}{$^a$IAU-MPC code: N50, HFOSC Resolution: 0.296 arcsec/pixel.}\\
\multicolumn{11}{|l|}{$^b$EMPOL Resolution: 0.720 arcsec/pixel.}\\
\multicolumn{11}{|l|}{$^c$IAU-MPC code: 954 , Camelot-2 Resolution: 0.336 arcsec/pixel.}\\
\multicolumn{11}{|l|}{$^d$LISA Resolution: 0.326 arcsec/pixel.}\\
\hline
\end{tabular}
\label{observations}
\end{scriptsize}
\end{table}
\normalsize
\subsection{Spectroscopy}
\begin{figure}
   \centering
   \includegraphics[width=0.7\linewidth]{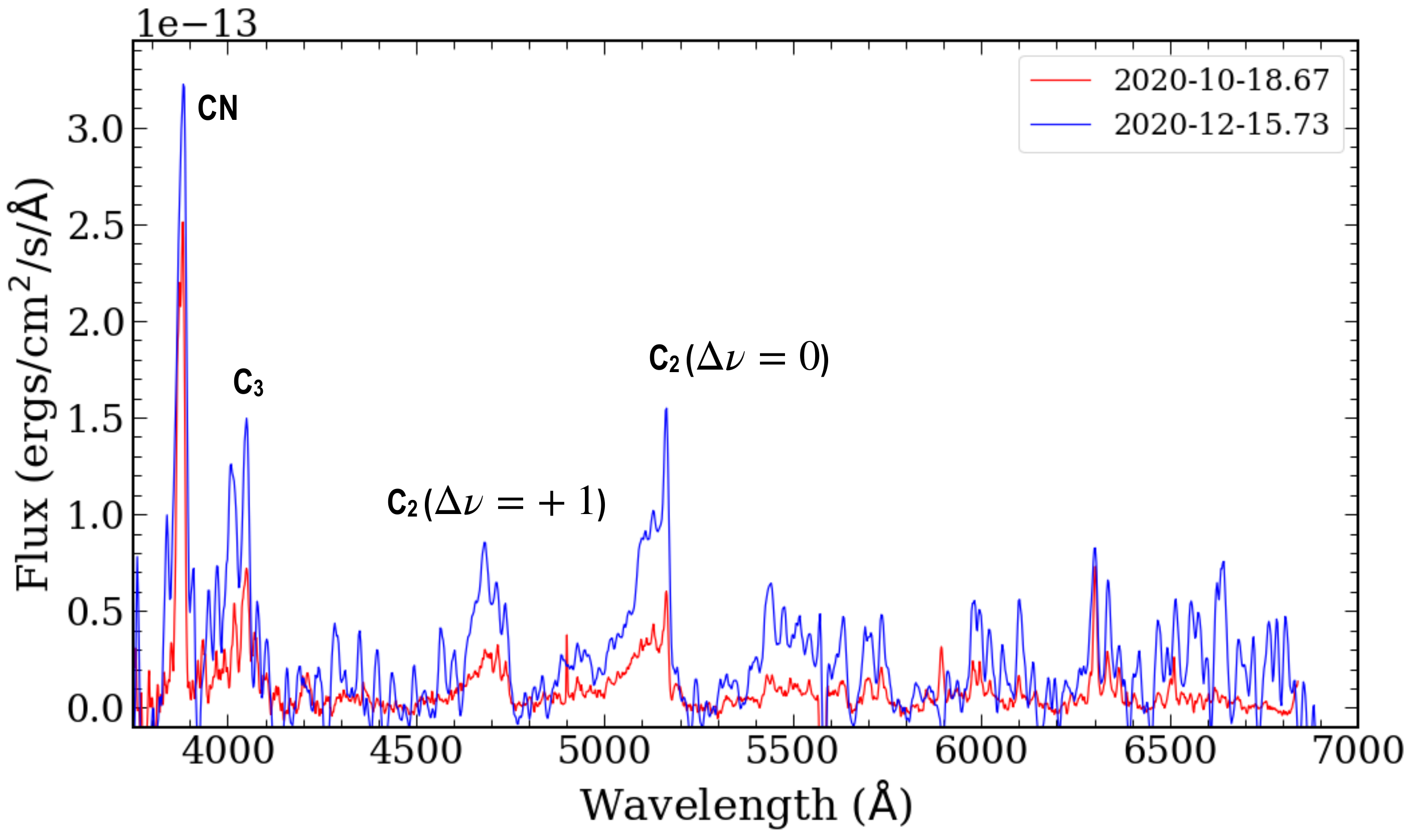}
   \caption{\textit{Red:} Optical spectrum of 156P observed pre-perihelion with the HFOSC instrument on HCT on 2020-10-18.67 UT; \textit{Blue:} Optical spectrum of 156P post-perihelion observed with the LISA instrument on MIRO on 2020-12-15.73 UT. }
              \label{spectra}
    \end{figure}

Considering the perihelion date of the comet, 2020 November 17, we have covered two epochs in spectroscopy, pre-perihelion and post-perihelion. Even though the two epochs are pre and post-perihelion, it is interesting to observe that the heliocentric distance at the time of observations are similar for both.\\ 
During the first epoch, the comet was observed from HCT using the HFOSC instrument. The instrument details and the slits used are as described by \citet{borisov_aravind}. Grism 7 providing a wavelength range 3800-6840 \AA, along with a long slit of length 11 arcmin and width 1.92 arcsec providing a resolving power of 1330, was used for observations. The comet and a separate sky frame were obtained with an exposure of 1200 seconds each. Standard star, \textsc{BD+28 4211}, from the catalog of spectroscopic standards in \textsc{IRAF} was observed for flux calibration using a slit of width 15.41 arcsec in order to avoid light loss. Halogen lamp spectra, zero exposure frames and FeAr lamp spectra were obtained for flat fielding, bias subtraction and wavelength calibration respectively.\\
During the second epoch, the comet was observed from MIRO with the help of the Long slit Intermediate resolution Spectrograph for Astronomy (LISA). The details regarding  the instrument is given by \citet{Kumar_thesis}. The instrument provides an effective wavelength range of 3800-7000 \AA~at a resolving power of 1000. A long slit, oriented in the N-S direction, 3.36 arcmin in length and 1.76 arcsec width was used for the observation of both comet and standard. In this case, due to the narrow slit, there will be loss of flux from the standard star resulting in an underestimation in the flux of the comet while performing flux calibration. Hence, a slit correction factor, as explained in \citet{slit_correction}, is introduced while extracting the spectrum. The comet and a separate sky frame were obtained with an exposure of 900 seconds each. During this epoch, standard star \textsc{HD 74721} was observed for flux calibration. Tungsten lamp spectra, zero exposure frames and ArNe lamp spectra were obtained for flat fielding, bias subtraction and wavelength calibration respectively.  
During both epochs, solar analog HD 81809 (G2V) \citep{HB} was observed in order to remove the continuum from the comet spectrum. \\
The comet was observed in non-sidereal tracking mode available in both facilities. In both  epochs, the sky frame was obtained 1$^\circ$ away from the photocentre of the comet in order to avoid any possible contribution from cometary emissions. The basic reductions, cosmic ray correction, sky subtraction and calibrations for all the frames were performed using standard IRAF packages. \\
The spectrum was extracted from the reduced comet frames observed on two epochs with the help of scripted PYTHON routines. The standard star spectrum was used to obtain the characteristic trace of the spectrum for the instrument. The same trace, with necessary corrections, was used to trace the comet spectrum along the dispersion axis and thus extract its spectrum along the spatial axis. The standard star spectrum was extracted using the IRAF \textit{apall} task, since it allows to perform an effective sky subtraction using regions on both sides of the target. Proper wavelength calibration and flux calibration were performed using standard IRAF packages. An appropriately scaled and slope corrected Solar spectrum (see \ref{continuum_removal})  was used to remove the contribution from the continuum.  The flux calibrated spectra of the comet observed on both epochs, after continuum removal, is  shown in Fig. \ref{spectra}.\\  

\subsection{Imaging}
The comet 156P was observed in imaging mode from both HCT and IAC-80 telescopes across the months 2020 October to December. As described in table \ref{observations}, the comet was observed from HCT on 2020 October 18 in the broadband \textit{R} filter. The Landolt's standard star field \citep{landolt} ru 149 in filter \textit{R}, twilight flat and bias frames were also acquired so as to perform the standard photometric analysis. The reduction and photometric analysis of the images were performed as described in \citet{borisov_aravind}. During the rest of the imaging epochs using the IAC-80 telescope, the comet was observed under the Pro-Am Cometary Morphological Evolution Study (COMES) project\footnote{\url{https://cometografia.es/comes/imagenes/}} using the CAMELOT-2\footnote{\url{http://research.iac.es/OOCC/iac-managed-telescopes/iac80/camelot2-2/}} camera containing a 4K $\times$ 4K pixels CCD with a resolution of 0.336 arcsec per pixel. Dark, flat and bias frames were also obtained during all the epochs in order to perform the standard reductions. Firstly,  the standard reduction was performed using Python subroutines. Then, the object images with the worst FWHM and deviation from median background level were rejected using PixInsight software. Finally, all the images were aligned to stars, then aligned to comet using the Maxim-DL software and then integrated with the average method.
\begin{figure}[h]
   \centering
   \includegraphics[width=0.9\linewidth]{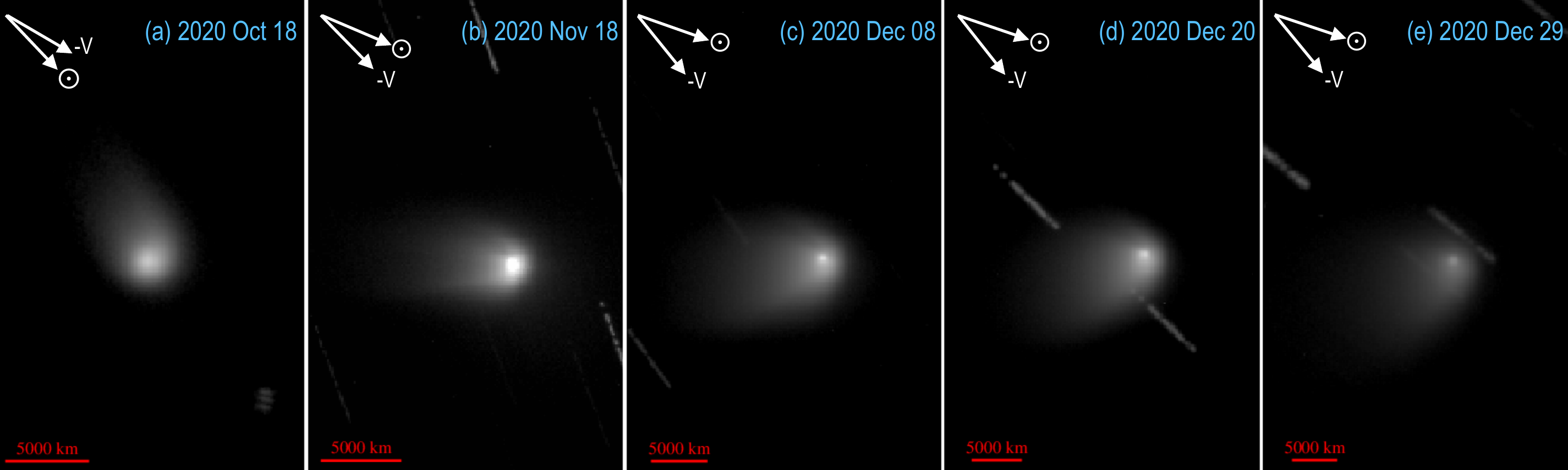}
   \caption{Compilation of few of the reduced imaging data obtained for comet 156P/Russell-LINEAR. North is up and East is on left for all the panels and the red solid line depicts 5000 km on the sky at the distance of the comet. The date of observation, direction to the Sun and the direction of negative heliocentric velocity are also mentioned in all the panels. Panel \textit{(a)} depicts observation from HCT R filter and the rest are from IAC observed in the absence of any filter. All the images have been scaled for the best visualization and the resulting gray scale does not represent their actual magnitude.} 
              \label{original}
    \end{figure}
    
\begin{figure}[h]
   \centering
   \includegraphics[width=0.9\linewidth]{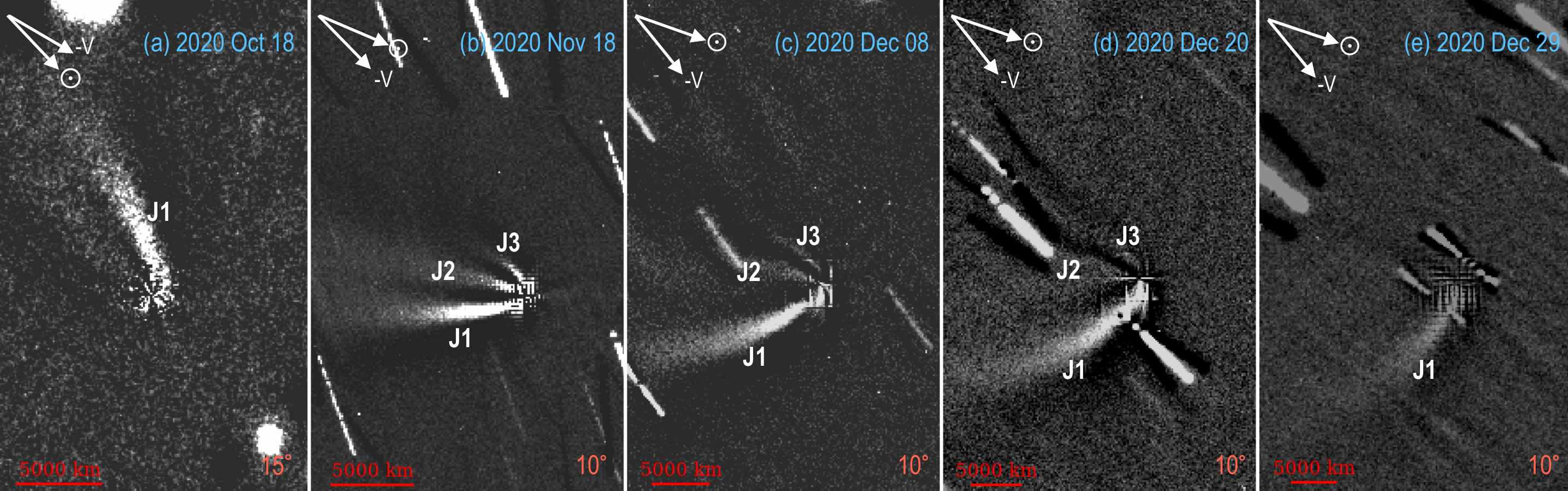}
   \caption{Compilation of LS processed outputs of imaging data of multiple epochs displayed in Fig. \ref{original}. North is upwards and East is towards the left for all the panels. The angle of rotation employed for the technique is mentioned in orange coloured text on the bottom right of each panel. The red solid line depicts 5000 km on the sky at the distance of the comet. The date of observation, direction to the Sun and the direction of the negative heliocentric velocity vector are also mentioned in all the panels.}
              \label{ls_processed}
    \end{figure}
    
Fig. \ref{original} compiles few of the imaging observations carried out for comet 156P. To an extent, it illustrates the variation in strength of the coma as the comet crossed perihelion and moved away from the Sun. At a glance, most of the  images reveals the presence of probable dust jets arising from the comet nucleus. Further analysis and detailing of these features are described in Section \ref{morphology}.

\subsection{Polarization}\label{pol_obs}
\begin{figure}[h]
\begin{subfigure}{0.49\linewidth}
\centering
\includegraphics[width = 0.85\linewidth]{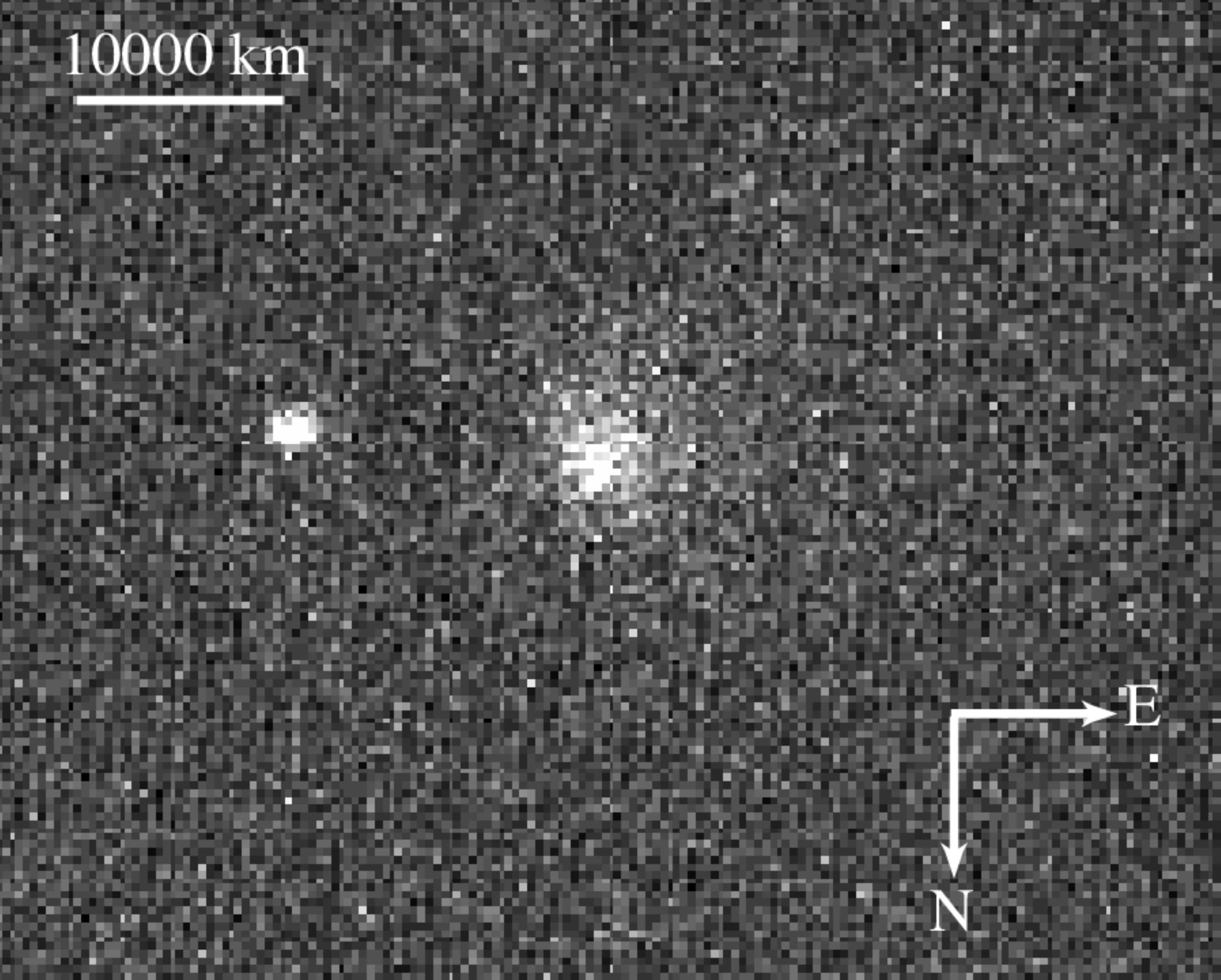}
\subcaption{{Single raw frame (0.5 s exposure) of the comet obtained during the polarization observation. Gain is used appropriately based on the comet brightness.}}
\label{single}
\end{subfigure}\hfill
\begin{subfigure}{0.49\linewidth}
\centering
\includegraphics[width = 0.8\linewidth]{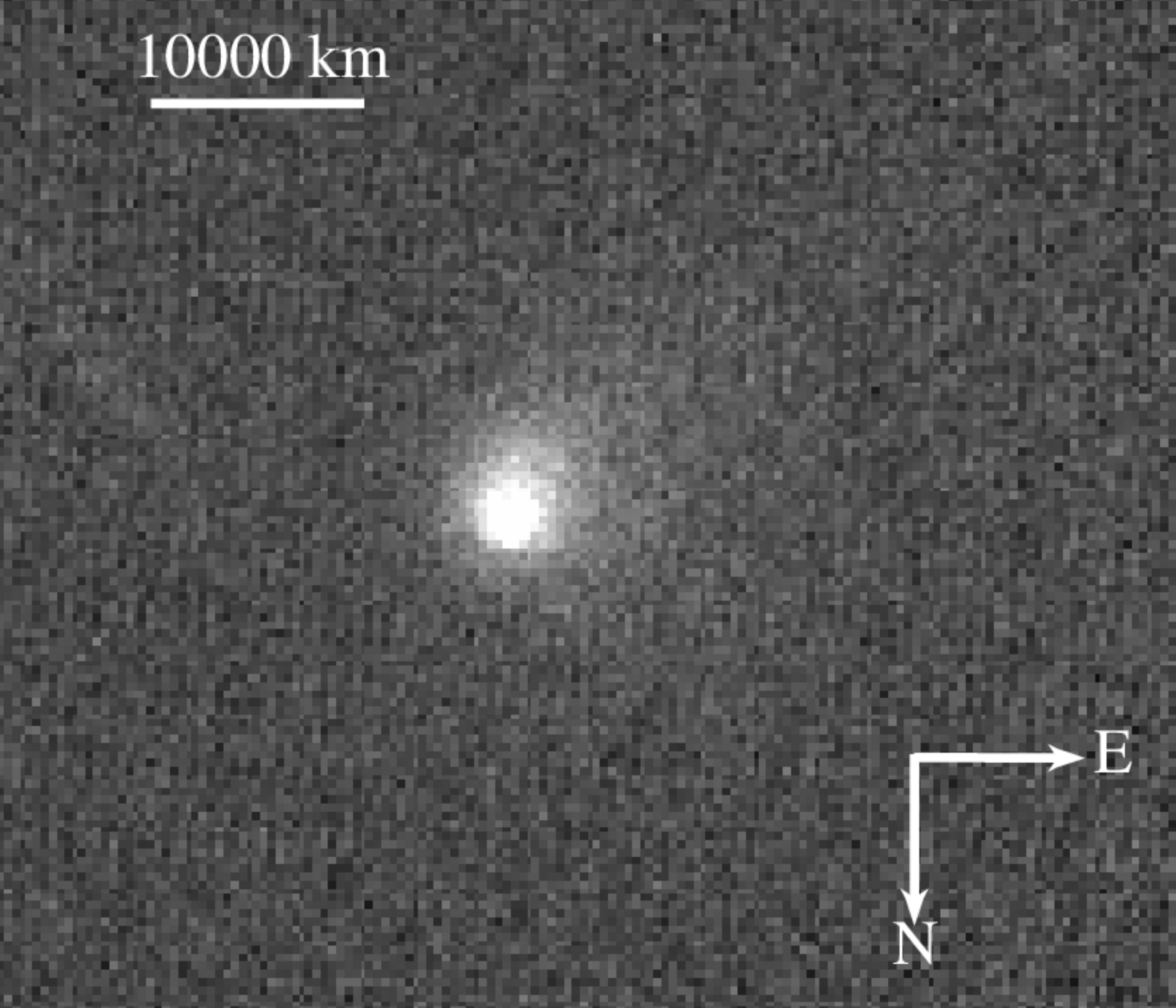}
\subcaption{{One frame among the 48 final frames where the corresponding single frames have been reduced, shifted and combined to increase the SNR. }}
\label{combined}
\end{subfigure}
\caption{Comparison of the raw output of EMPOL instrument with the reduced and combined image used for analysis as observed on 2020-12-16. }
\end{figure}

The 1.2 m MIRO telescope houses an EMCCD based optical polarimeter. It is an in house developed instrument \citep{empol} with a 1K $\times$ 1K ANDOR EMCCD having a plate scale of 0.72 arsec per pixel in a 4 $\times$ 4 binning mode.  The polarization measurement is carried out using a rotating half-wave plate (48 steps per rotation) as the modulator and a wire grid polarizer as the analyser. A filter wheel with the Sloan filters \textit{u, g, r, i, z} is placed after the analyzer. The comet 156P was observed at two different phase angles, 38.9$^\circ$ and 42.34$^\circ$, in the \textit{Sloan i} filter in order to make sure that the light observed was solely the scattered
light from the dust present in the coma. The total number of frames to be acquired was decided based on the effective exposure required per frame, after combining the respective angles. Considering the exposure per frame is 0.5 s, a complete rotation of the half wave plate consists of 48 frames and an effective exposure of 10 s per frame was required, a total of 1008 frames were acquired. The raw frame corresponding to one of the 48 steps is as shown in Fig. \ref{single}. Another set of data, namely sky frames, using the same settings was taken with the telescope pointed about 1$^\circ$ away from the comet. These would be used to remove any contribution of sky polarization in the comet frames. Twilight flat frames and bias frames were also acquired as per requirement. On both the epochs, out of a few high polarised standards listed in \citet{pol_standard}, BD+64 106 and HD 25443 were observed in order to confirm the operation of the instrument and to obtain the zero point correction of the polarization angle. Both the standards were observed in \textit{sloan i} filter with an effective exposure of 1 s per frame. During the reduction of both comet and standard star frames, all the single frames were initially bias subtracted and flat fielded. Since there would be slight shift in the object position from frame to frame considering the amount of time for which the frames are acquired, they were shifted to a common point and then median combined in order to build up the signal. After the reduction process we are left with 48 frames, each with 10 s effective exposure, (see Fig. \ref{combined}) which would now be used to compute the degree of polarization and polarization angle as described in the Section \ref{pol_anal}.

\section{Data Analysis}\label{analysis}

\subsection{Spectroscopy}
\subsubsection{Gas production rates}\label{column}
 Strong emissions from $CN (\Delta \nu = 0)$, $C_3 (\lambda4050$\AA), $C_2 (\Delta \nu = +1)$ and $C_2 (\Delta \nu = 0)$ can be observed during both the epochs. Comparing the pre and post-perihelion spectra (see Fig. \ref{spectra}), it can be inferred that there has been an increase in the activity, post-perihelion, even though the heliocentric distance was almost the same during both the epochs. In order to compute the production rates of the various molecules detected, the spectra were extracted to one side of the spatial axis with equal aperture size at different distances from the photocentre. The formula,
 \begin{equation}
     N (y) = \frac{4\pi}{g}\frac{F(y)}{\Omega},
     \label{column_density}
 \end{equation}
 gives the column density (molecules per centimetre cube) corresponding to the flux (\textit{F}) within the wavelength range of the molecule (defined by \citet{langland-shula}) extracted from an aperture, at a distance y from the photocentre, subtending a solid angle $\Omega$. The Solid angle, in steradians, is computed as the product of the slit width and the aperture size in radians.  Here, \textit{g} is the fluorescence efficiency (ergs per molecule per second). \textit{g} at 1 AU for molecules $C_2$ and $C_3$ were taken from \citet{Ahearn_85} and were scaled to $r_h^{-2}$ in order to obtain the appropriate values to be used at the corresponding heliocentric distance ($r_h$). \citet{schleicher_CN_2010} have tabulated the g-factor of CN for different heliocentric distances and velocities. A double interpolation was performed on the provided table to obtain the exact \textit{g} values for the corresponding heliocentric distance and velocity at the time of observation. \citet{langland-shula} and \citet{kumar} have detailed a method to fit the Haser model \citep{haser} to the observed column density profile in order to compute the production rate (molecules per second) of the corresponding molecule. The production rate \textit{Q}, in molecules per second, is estimated using the minimum chi-square estimation between the observed column density and the theoretical column density computed using the equation,

\begin{table}[h]
\centering
\begin{scriptsize}
\renewcommand{\arraystretch}{1.6}
\caption{Scale lengths and fluorescence efficiency of different
molecules at both epochs of observation.}
\setlength\tabcolsep{5pt}
\begin{tabular}{| c |c |c | c  | c | c | c| c|}
\hline
     & \multicolumn{3}{|c|}{2020-10-18} & \multicolumn{3}{|c|}{2020-12-15}  \\
     \cline{2-7}
          & g-factor  & l$_p$  &  l$_d$   & g-factor  & l$_p$  &  l$_d$   \\ 
          Molecule & ($ergs mol^{-1} s^{-1}$) & (10$^4$ km) & (10$^4$ km) & ($ergs mol^{-1} s^{-1}$) & (10$^4$ km) & (10$^4$ km) \\\hline
         CN(0-0)&1.89$\times 10^{-13}$&2.47&39.99&2.36$\times 10^{-13}$ &2.44&39.94\\
         C$_2(\Delta \nu=0)$&2.36$\times 10^{-13}$&4.18&12.56& 2.40$\times 10^{-13}$ &4.13&12.38\\
         C$_3$&5.25$\times 10^{-13}$&0.53&5.14& 5.33$\times 10^{-13}$&0.52&5.06
   
              \\ \hline
\end{tabular}
\label{parameters}
\end{scriptsize}

\end{table}

 \small
 \begin{equation}
     N(y)=\frac{Q}{2 \pi v_{\mathrm{flow }}} \frac{\beta_{0}}{\beta_{1}-\beta_{0}} \int_{0}^{\infty} \frac{1}{y^{2}+z^{2}}\left(e^{-\beta_{0} \sqrt{y^{2}+z^{2}}}-e^{-\beta_{1} \sqrt{y^{2}+z^{2}}}\right) d z,
 \end{equation}
\normalsize
Here, \textit{y} is the projected distance from the centre of the comet nucleus, $v_{\mathrm{flow }}$ is the outflow velocity in centimetre per second, z is in the line-of-sight direction, and $\beta_0$ and $\beta_1$ are the inverse of the parent and daughter molecule scale lengths in centimetres. The scale lengths of the parent (l$_p$) and daughter (l$_d$) molecules were taken from \citet{Ahearn_85} and were scaled to $r_h^{2}$. The values that have been used for \textit{g, l$_p$ and l$_d$ } at the respective epochs are given in table \ref{parameters}. During the computation of production rates, the most uncertain factor is the outflow velocity of gas and dust from the nucleus. There are various scaling laws which can be incorporated for the computation \citep{langland-shula, 2P_rosenbush, whipple_outflow}. This work adopts the outflow velocity relationship $v_{\mathrm{flow}}=
0.85/\sqrt{r_h}$ as described in \citet{cochran_30years}. \\
\begin{figure}[h]
   \centering
   \includegraphics[width=0.7\linewidth]{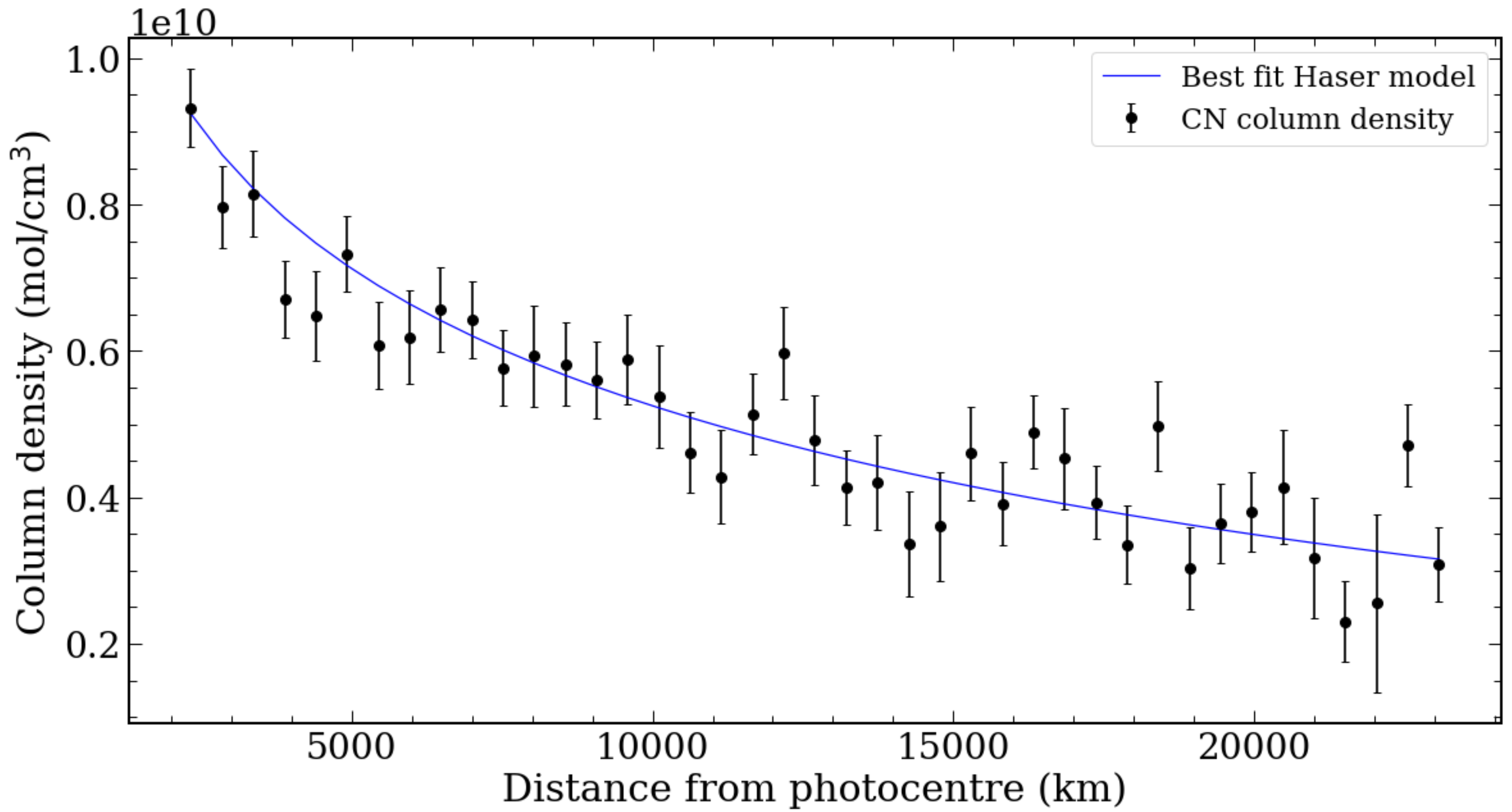}
   \caption{Column density profile of CN in the comet as observed on 2020-10-18. The blue solid line depicts the best Haser model fit. Error bars represent the standard error in column density at each respective points.}
              \label{CD}
    \end{figure}
Fig. \ref{CD} illustrates the observed column density profile of CN, plotted along with the best fit Haser model used to compute the
\begin{table}[h]
\centering
\setlength\tabcolsep{5pt}
\begin{scriptsize}
\renewcommand{\arraystretch}{1.6}
\caption{{Gas production rates of comet 156P at different epochs}}
\begin{tabular}{|r|c|r|r|r|r|r|c|c|}
\hline
 \multicolumn{1}{|c|}{Date} &  \multicolumn{1}{|c|}{Exposure} &\multicolumn{1}{|c|}{r{$_{H}$}}   & \multicolumn{1}{|c|}{$\Delta$} & \multicolumn{3}{|c|}{Production Rate (molecules per sec)}  & \multicolumn{1}{|c|}{Production rate ratio}&\multicolumn{1}{|c|}{Dust to gas ratio$^a$}\\ 

 \multicolumn{1}{|c|}{(UT)} & \multicolumn{1}{|c|}{(s)}&   \multicolumn{1}{|c|}{(AU)} & \multicolumn{1}{|c|}{(AU)} & \multicolumn{1}{|c|}{CN} & \multicolumn{1}{|c|}{C{$_{2}$}({$\Delta \nu = 0$})} & \multicolumn{1}{|c|}{C{$_{3}$}} & \multicolumn{1}{|c|}{Q(C$_2$)/Q(CN)} & \multicolumn{1}{|c|}{log[($Af\rho$)$_{BC}$/Q(CN)]}\\ 
  & &  &  & \multicolumn{1}{|c|}{$\times 10^{24}$}  & \multicolumn{1}{|c|}{$\times 10^{24}$}  &\multicolumn{1}{|c|}{$\times 10^{24}$}  &  &\\
\hline  
 2020-10-18.67 & 1200 & 1.38 & 0.48 & 3.06 $\pm$ 0.12 & 3.32 $\pm$ 0.41  & 0.51 $\pm$ 0.05 & 1.08 $\pm$ 0.17  & -22.48 $\pm$ 0.29 \\
 2020-12-15.73 & 900 &  1.37 & 0.67 & 9.43 $\pm$ 0.16 & 12.5 $\pm$ 0.9 & 2.32 $\pm$ 0.26  & 1.32 $\pm$ 0.12& -22.24 $\pm$ 0.16 \\ 
\hline 
\multicolumn{9}{|l|}{$^a$ $Af\rho$ corresponding to an aperture size of 10,000 Km has been used to compute the ratio}\\
\hline
\end{tabular}
\label{result_table}
\end{scriptsize}

\end{table} 
\noindent production rate as mentioned in table \ref{result_table}. The production rates of other detected molecules, $C_2$ and $C_3$, were also computed in similar manner for both epochs. The error obtained from the Haser model fitting is taken as the standard error in production rate while the standard errors in the column density are obtained from the propagation of errors in the parameters mentioned in Equation.\ref{column_density}. The error in flux is computed from the continuum region close to the bandpass of the molecule of interest.

 \subsubsection{Dust production}\label{dust}
\citet{Ahearn_Bowel_slope} defined a parameter $Af\rho$, which is a proxy to the amount of dust produced. The parameter is directly measurable using the formula,
\begin{equation}
    Af\rho = \frac{(2\Delta r)^2}{\rho}\frac{F_{\mathrm{comet}}}{F_{\mathrm{Sun}}}\frac{1}{S(\theta)},
    \label{afrho}
\end{equation}
where $\Delta$ is the geocentric distance in kilometres, $\rho$ the aperture size in kilometres, \textit{r} is the heliocentric distance in AU, and $F_{\mathrm{comet}}$ is the observed cometary flux within the bandpass of the filter. $F_{\mathrm{Sun}}$ is the incident Solar flux at 1 AU integrated over the bandpass of the filter and $S(\theta)$ is the phase function\footnote{Composite Dust Phase Function for Comets:  \url{https://asteroid.lowell.edu/comet/dustphaseHM_table.txt}} corresponding to the phase angle at the time of observation, as defined in \citet{schleicher_phasefunction}. In the current work we have computed $Af\rho$ in the blue continuum (4390–4510 \AA) (hereby called BC) and green continuum (5200–5320 \AA) (hereby called GC) narrow-band filters using the spectroscopic data of both the epochs and in the Johnson-Cousins \textit{R} filter using an imaging data obtained on the first epoch using HCT. The specifications of the narrow band filters are discussed in \citet{HB}.

While using the spectroscopic data to compute $Af\rho$, it is necessary to convert the observed flux obtained using long slit into a full disk flux. To attain this, we have used a geometrical conversion factor, in a similar way as used by \citet{langland-shula}, for each column of fixed aperture moving outwards the centre of the nucleus. The geometrical conversion factor is defined as the ratio of area of the circular/annular area to the area of the aperture used. The spectrum extracted for each column of fixed aperture, as explained in Section \ref{column}, is now multiplied with the corresponding conversion factor to convert the spectrum into circular/annular flux. In order to obtain the full disk flux within each increasing aperture radius, the flux computed for each circular/annular region within the required radius are added up. Now, to obtain the cometary flux within the band pass of the narrow band filter, the spectrum of the comet was initially convolved with the transmission profile of the filter re-sampled to match the resolution of the instrument. Finally, the resulting spectrum was averaged over the bandpass of the narrow band filter as given by the equation 
\begin{equation}
    F_{\mathrm{comet}} = \frac{\int_{\lambda_1}^{\lambda_2} I(\lambda)S(\lambda)d\lambda}{\int_{\lambda_1}^{\lambda_2} S(\lambda)d\lambda}. 
\end{equation}

Here, $\lambda_1$ and $\lambda_2$ are the wavelength range of the bandpass filter used, I($\lambda$) is the flux as a function of wavelength and S($\lambda$) is the transmission profile of the filter. The solar flux at 1 AU for the respective filter was obtained by using the magnitudes of the solar analogue stars given in \citet{HB}.

\begin{figure}
\centering
\subcaptionbox{Pre-perihelion (2020-10-18) Af$\rho$ profiles of the comet computed for broadband R filter and narrow band blue and green continuum filters. R filter profile is obtained from the imaging data and the narrow band filter profiles are obtained using the spectroscopic data. The standard error in the measurements are 23 cms. \label{afrho_hct}}%
  [0.49\linewidth]{\includegraphics[width = 1\linewidth]{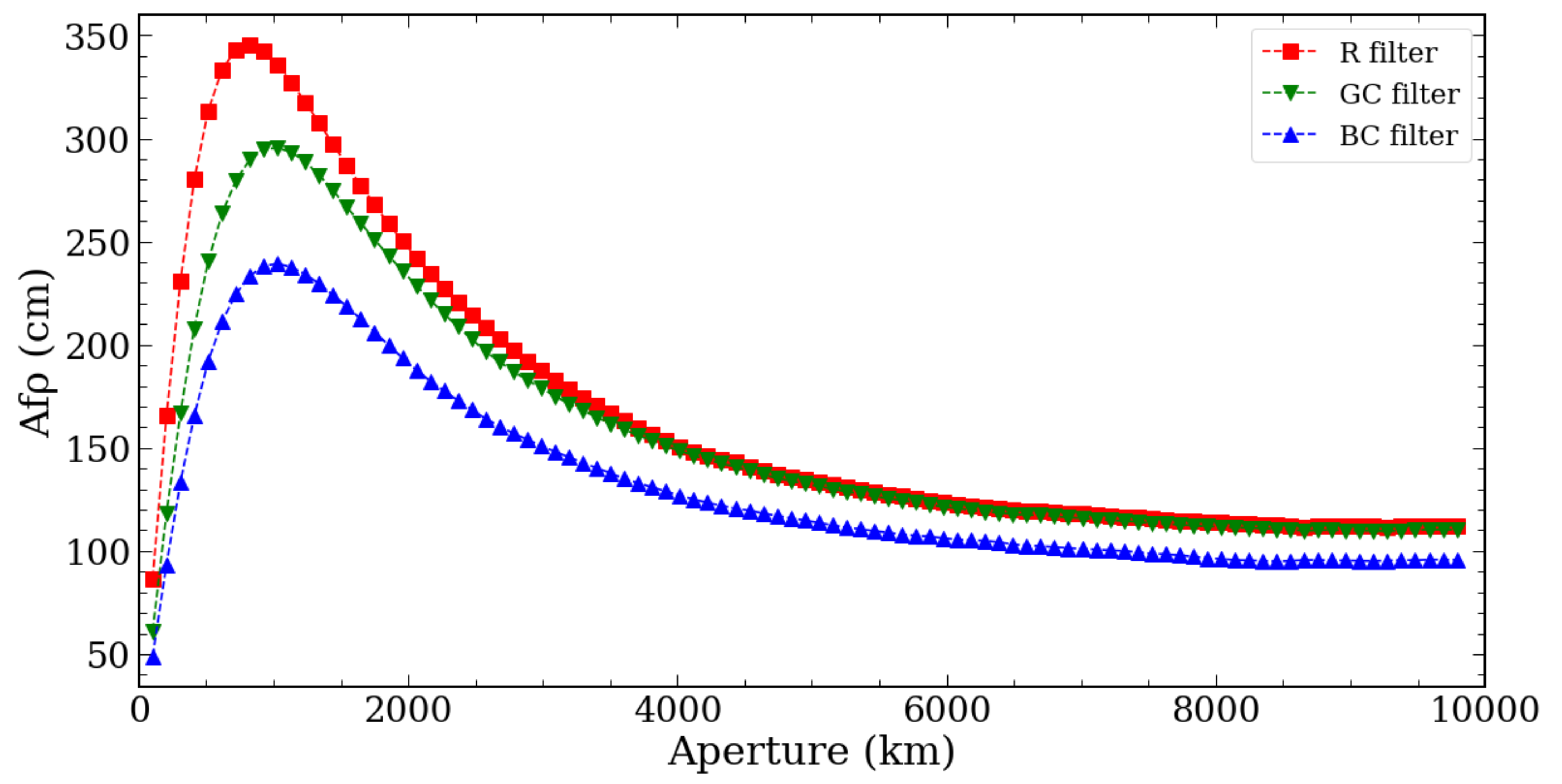}}
  \hfill
\subcaptionbox{Post-perihelion (2020-12-15) Af$\rho$ profiles of the comet in narrow band continuum filters. The spectroscopic data has been used to obtain the profiles. The standard error in the measurements are 81 cms. \label{afrho_lisa}}
  [0.49\linewidth]{\includegraphics[width = 1.02\linewidth]{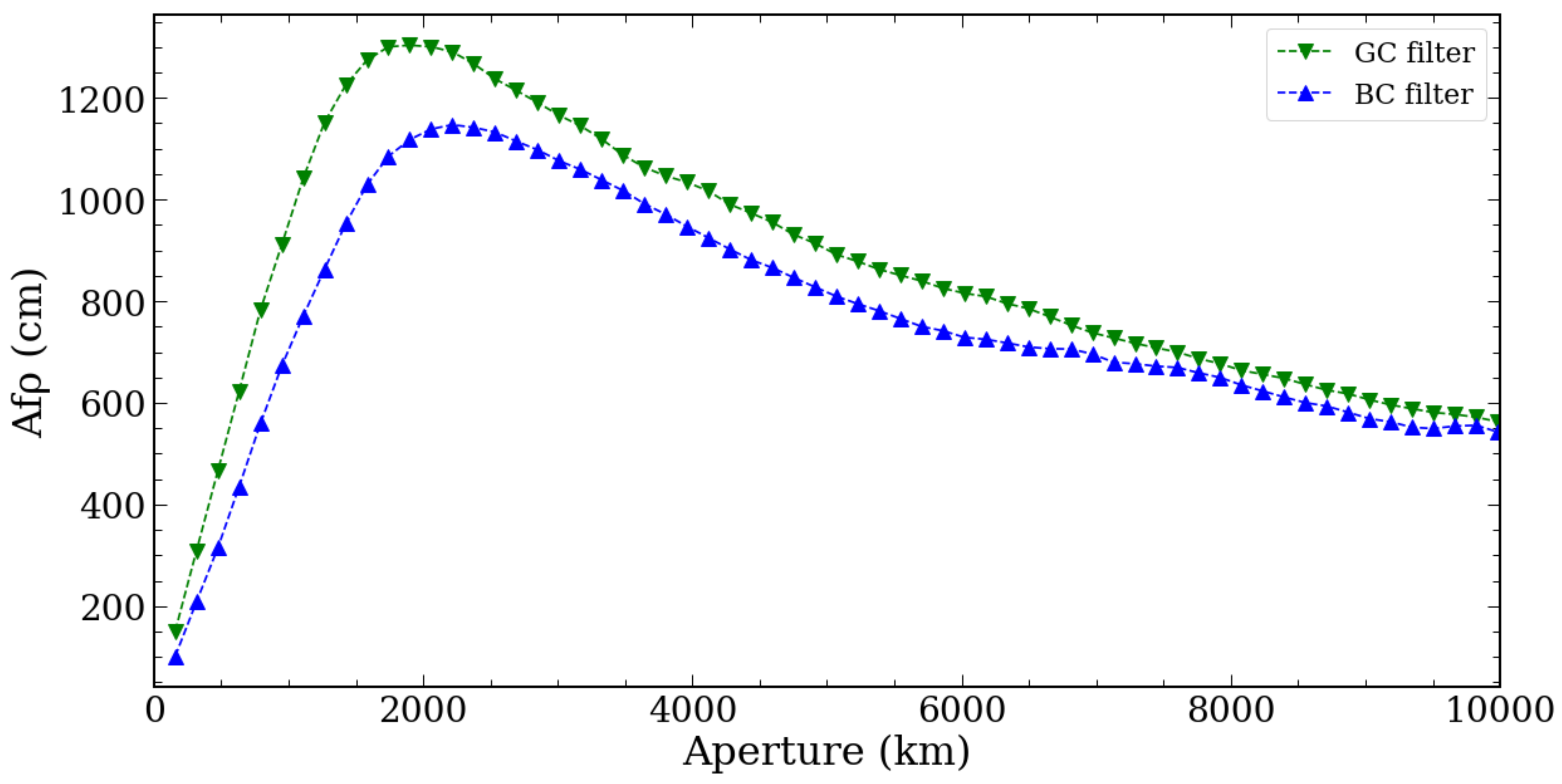}}
\caption{Pre and post perihelion Af$\rho$ profiles in various filters.}
\end{figure}

\begin{figure}[h]
  \centering
  \includegraphics[width=0.7\linewidth]{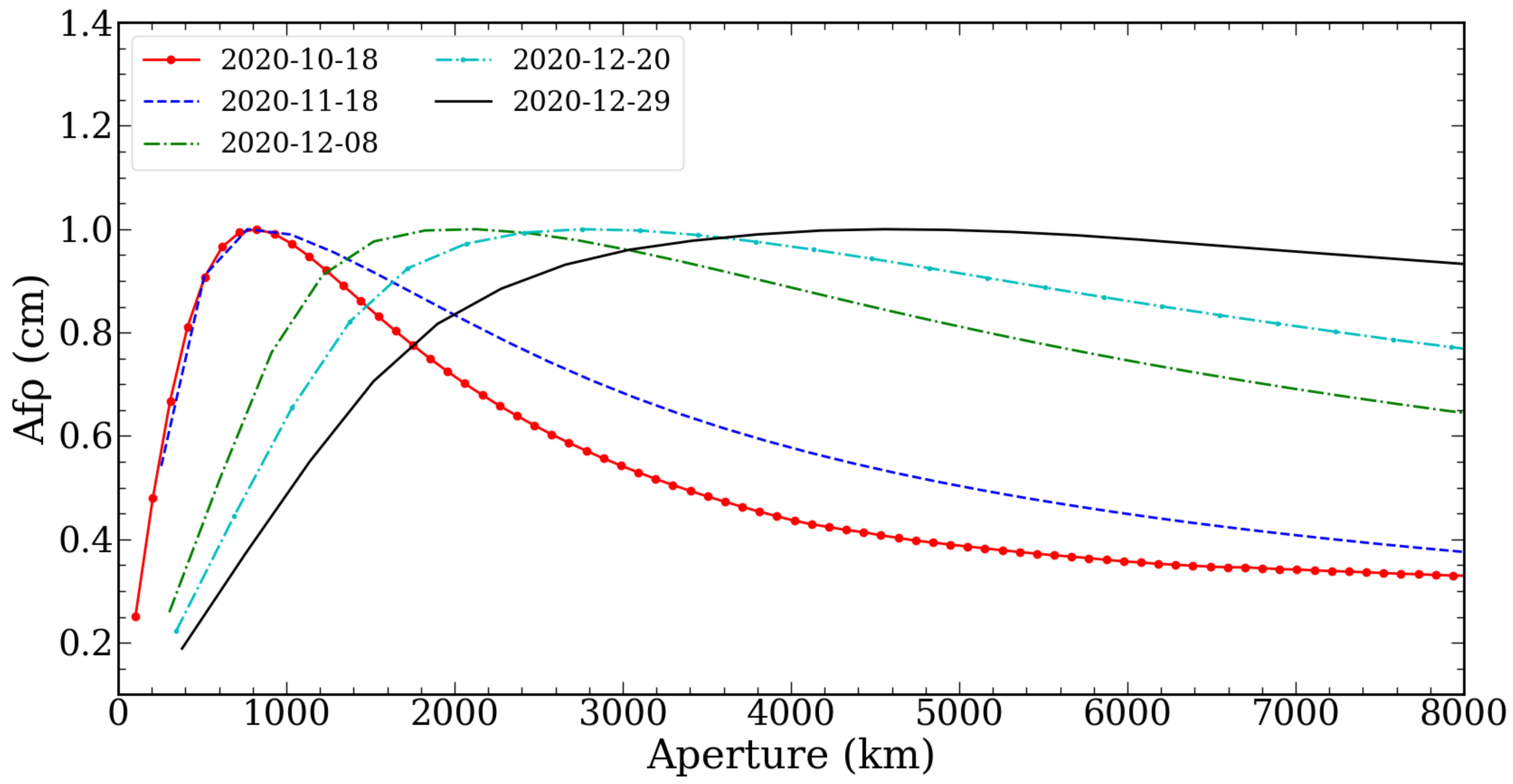}
  \caption{Comparison of the normalized Af$\rho$ profiles derived from a few selected imaging data observed using the HCT and IAC-80 telescope. The profile represented by the black solid line closely resembles the Af$\rho$ profile of a comet with a steady state outflow of dust.}
              \label{afrho_alldate}
    \end{figure}

While using the imaging data in \textit{R} filter, aperture photometry is performed to compute the apparent magnitude of the comet, as explained in \citet{borisov_aravind}, which is then used to determine the $Af\rho$. Apertures of increasing radius can be chosen to obtain the profile of $Af\rho$ similar to what is obtained while using spectroscopic data. 
Fig. \ref{afrho_hct} and \ref{afrho_lisa} represents the $Af\rho$ profile of the comet in various filters as observed on 2020 October 18 and 2020 December 15. Standard error in Af$\rho$ is computed by the propagation of errors present in the parameters mentioned in Equation.\ref{afrho}.\\
Due to the absence of proper calibration standard stars during the imaging epochs of 156P obtained from IAC-80, the observed data was used to analyse any variation in Af$\rho$ profile over the period. Even though the observations are made in the absence of any filter, the contribution from the molecular emissions to the Af$\rho$ profile would be very less considering the physical scale at which the variations are seen. Also, the similarity of the Af$\rho$ profiles obtained on 2020-10-18 and 2020-12-15, derived for wavelength range corresponding to proper dust emissions, with those obtained on 2020-11-18 and 2020-12-08 in clear filter also suggests that the profile characteristics are dominated by the dust present in the coma. Hence, as shown in Fig. \ref{afrho_alldate}, the profiles do not indicate the actual numbers of Af$\rho$, rather are normalised to their peak values in order to compare the variation in the characteristics of the profile across the dates. In addition, the dominance of dust in the comet, especially in the inner coma, is explicitly seen from Fig. \ref{continuum_fitted} in \ref{continuum_removal} where the solar continuum is matching the comet spectrum (extracted at the photocentre with an aperture of 103 km) over the entire wavelength range without any sign of emissions from other molecules. The gas emissions from different molecules starts to dominate only when the spectrum is extracted for larger apertures.

\subsection{Polarization}\label{pol_anal}
\begin{figure}
   \centering
   \includegraphics[width=0.7\linewidth]{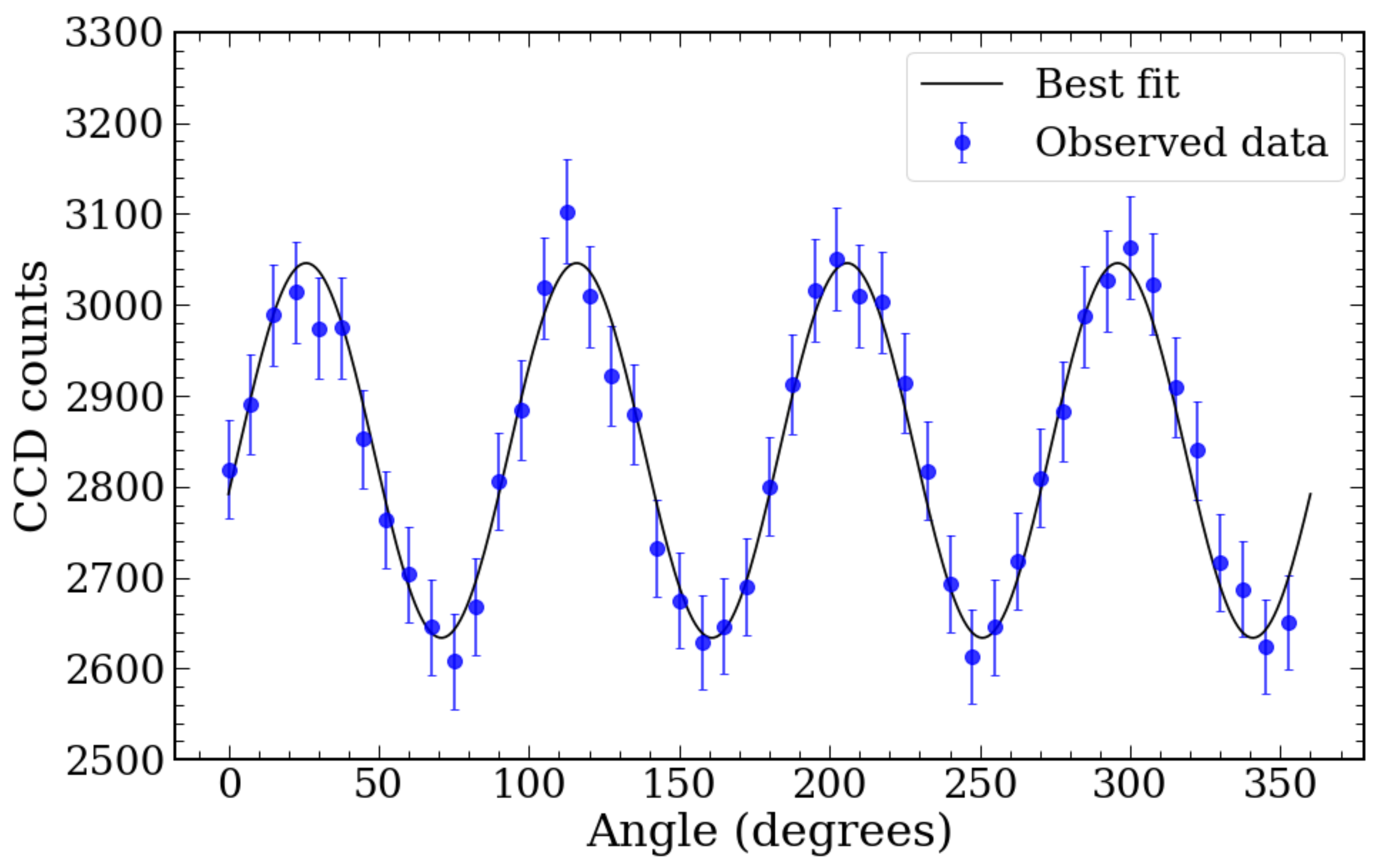}
   \caption{Intensity modulation derived from the 48 final frames obtained from the polarization observation of the comet. The black solid line portray the best fit line for the equation.\ref{fit_equation} used to compute Stokes' parameters. The error bar represents the standard error in photometry at each angle}
              \label{pol_mod}
    \end{figure}

As mentioned in Section \ref{pol_obs}, the final product after all the reduction process would be 48 frames of the standard star or comet which is to be used to compute the degree of polarization and polarization angle. The 48 frames correspond to 48 angles spanning from 0 to 360 degree. A PYTHON code is used to perform aperture photometry on all the frames of the object being analysed to obtain the counts within a fixed aperture. Knowing the counts corresponding to each of the 48 angles, the equation
\begin{equation}\label{fit_equation}
    I_o = \frac{1}{2}[{I+Q\cos{4\theta}+U\sin{4\theta}}],
\end{equation}
where \textit{I$_o$} is the observed intensity corresponding to the $\theta$ in radians, is used to get the best fit Stokes' parameters \textit{I, Q, U}. Fig. \ref{pol_mod} illustrates the modulation of the observed intensity across the 48 frames of the comet 156P, over plotted with the best fit curve used to determine the Stokes' parameters. Once the Stokes' parameters are derived, the degree of polarization (DoP) and the position angle for the plane of polarization (PA) are calculated using the equations \ref{PF} and \ref{PA} respectively.

\begin{equation}\label{PF}
    DoP =\frac{\sqrt{Q^{2}+U^{2}}}{I}
\end{equation}

\begin{equation}\label{PA}
    PA =\frac{1}{2} \tan ^{-1}\left(\frac{U}{Q}\right)
\end{equation}

The uncertainties in the measured quantities are determined using error propagation \citep[eg.][]{error_polarisation}, making use of the uncertainties in the derived Stokes' parameters obtained from the curve fitting. These analysis techniques are performed on the comet and the standard star frames observed on both epochs. The observational results for the observed standards are as mentioned in table \ref{pol_stan_result}, where $p_{\textit{obs}}$ and $\theta_{\textit{obs}}$ depicts the directly measured degree of polarization and polarization angle, \textit{p$_0$} and $\theta_0$ represents the actual values taken from \citet{pol_standard} (for Johnson-Cousins \textit{I} filter) and $\theta_{\textit{off}}$ portrays the zero point offset in the polarization angle. The values of degree of polarization and polarization angle of the comet observed for an aperture of 10000 km are specified in table \ref{comet_pol_result}.\\
Radially increasing apertures were used to check for any spatial variation in the degree of polarization as observed in comet 67P \citep{67P_pol}. The observed degree of polarization for multiple increasing apertures were similar within error bars, ruling out the presence of any systematic changes in the physical properties of the dust particles present in the coma. Average of the zero point offset in the polarization angle, for each epoch, obtained from table \ref{pol_stan_result} is used to obtain the actual polarization angle of the comet ($\theta_\mathrm{corr})$. The observed polarization is found to be positive over the whole coma since the angle of linear polarization is observed to be perpendicular to the scattering plane.  

\begin{table*}
\centering
\setlength\tabcolsep{5pt}
\begin{scriptsize}
\renewcommand{\arraystretch}{1.6}
\caption{Observational results of polarized standard stars. }
\begin{tabular}{|c | c |c | c  | c | r | r | c |}
\hline
  Date & Star & Filter & \textit{p$_0$}  &  $\theta_0$ & \multicolumn{1}{|c|}{\textit{p}$_{\mathrm{obs}}$}  & \multicolumn{1}{|c|}{$\theta_{\mathrm{obs}}$} & $\theta_{\mathrm{off}}$=$\theta_{0}-\theta_{\mathrm{obs}}$ \\
           & & & (percent) & ($\circ$)     & \multicolumn{1}{|c|}{(percent)}   & \multicolumn{1}{|c|}{($\circ$)}    & ($\circ$) \\ \hline
  13-11-2020 & HD 25443 & \textit{i} & 4.249 $\pm$ 0.041 & 134.21 $\pm$ 0.28 & 4.12 $\pm$ 0.29  & 19 $\pm$ 2   &  115 $\pm$ 2  \\
   & BD +64 106 & \textit{i} & 4.696 $\pm$ 0.052 & 96.89 $\pm$ 0.32 & 4.60 $\pm$ 0.32 & -15 $\pm$ 3 & 112 $\pm$ 4
              \\ \hline
   16-12-2020      &  HD 25443 & \textit{i} & 4.249 $\pm$ 0.041 & 134.21 $\pm$ 0.28 & 3.97 $\pm$ 0.31 & 27 $\pm$ 4 & 107 $\pm$ 4\\
  &  BD +64 106   & \textit{i} & 4.696 $\pm$ 0.052  & 96.89 $\pm$ 0.32 & 4.76 $\pm$ 0.33 & -9 $\pm$ 2  &  106 $\pm$ 3 \\ \hline
\end{tabular}

\label{pol_stan_result}
\end{scriptsize}

\end{table*}

\begin{table}
\centering
\setlength\tabcolsep{5pt}
\begin{scriptsize}
\renewcommand{\arraystretch}{1.6}
\caption{Observational results of comet 156P. }
\setlength\tabcolsep{3pt}
\begin{tabular}{| c |c |c | c  | c | c |}
\hline
    Date & Filter & Phase Angle & \multicolumn{1}{|c|}{\textit{p}$_{\mathrm{obs}}$}  & \multicolumn{1}{|c|}{$\theta_{\mathrm{obs}}$} & $\theta_{\mathrm{corr}}$=$\theta_{obs}+\theta_{\mathrm{off(avg)}}^*$ \\
           &    & ($\circ$)& \multicolumn{1}{|c|}{(percent)}   & \multicolumn{1}{|c|}{($\circ$)}    & ($\circ$) \\ \hline
   13-11-2020 & \textit{i} & 38.9 &5.26 $\pm$ 0.39  & 27.86 $\pm$ 4.80   & 141 $\pm$ 5   \\
    16-12-2020 & \textit{i} & 42.3 & 6.80 $\pm$ 0.21 & 43.93 $\pm$ 1.10 & 151 $\pm$ 2
              \\ \hline
\multicolumn{6}{|l|}{$^*$ The average values of zero point offset in polarization angle for each epoch are}\\ \multicolumn{6}{|l|}{113.458 $\pm$ 2.07 and 106.73 $\pm$ 2.24 respectively.}\\
\hline
\end{tabular}
\label{comet_pol_result}
\end{scriptsize}

\end{table}

\subsection{Dust modelling}\label{model}
To model the observed polarization-phase data explained in Section \ref{pol_anal},  we consider the recent comet dust model introduced by \citet{pritish_modelling}, which incorporates  different morphology of dust particles in accordance with the findings from the \emph{Rosetta}/\textsc{Midas} and \emph{Rosetta}/\textsc{Cosima} instruments \citep{mannel_MIDAS, guttler_ROSETTA}. The model provides best fit results for both long period as well as short period comets and also explains the wavelength dependence of polarization in the narrow band filters. For the comet 156P we used a mixed morphology of dust particles consisting of hierarchical aggregates (HA) and agglomerated debris (Solids) having inhomogeneous mixture of silicate minerals and carbonaceous compounds under power-law size distribution. These are similar to those used to model the observed polarimetric-phase curve for the comet 67P/Churyumov–Gerasimenko (hereafter 67P) due to similar physical and dynamical characteristics. 
\begin{figure}[h]
	\centering
	\includegraphics[width=0.7\linewidth]{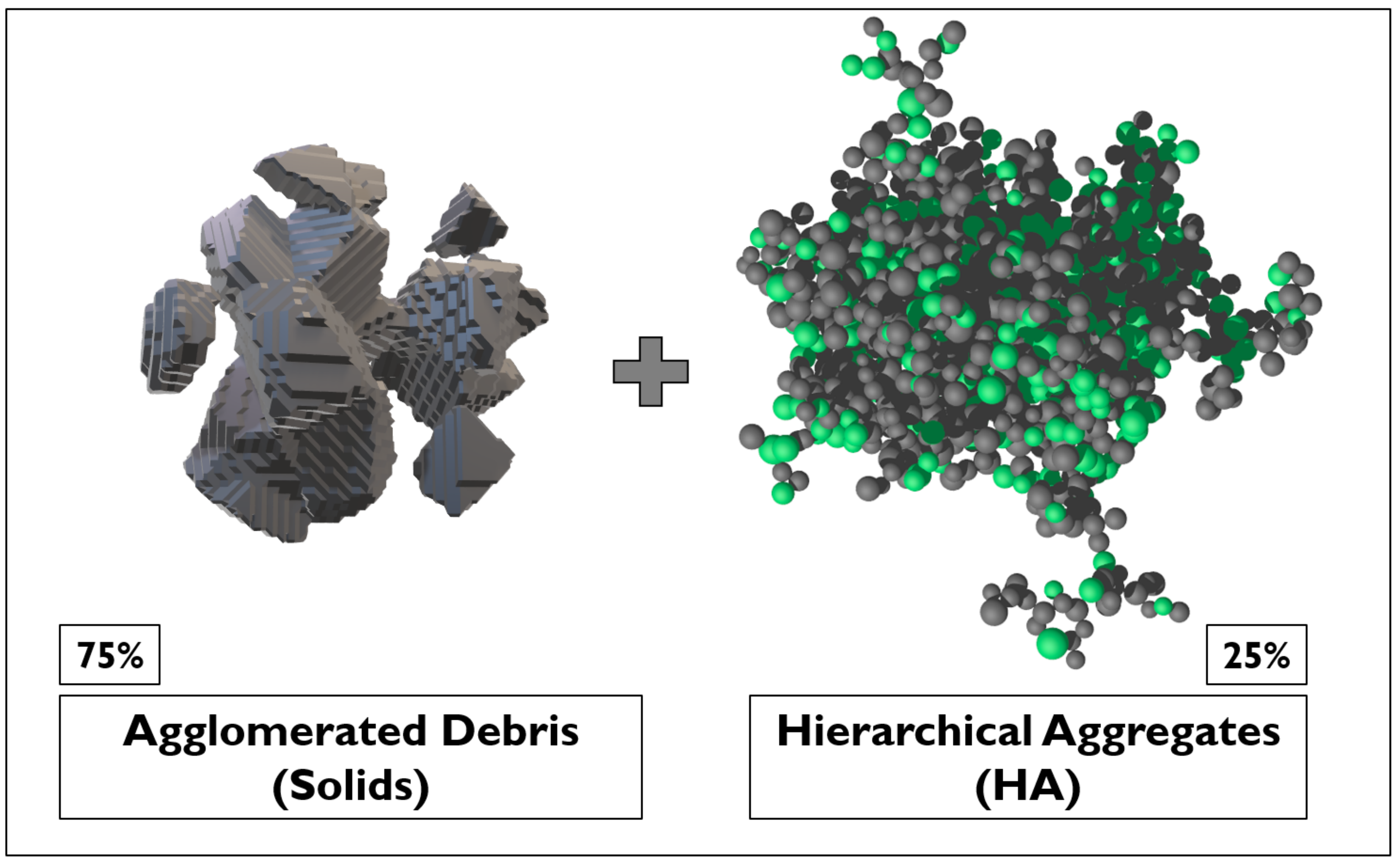}
    \caption{3D visualisation of the heterogeneous particle morphology and mixture used to model the observed polarization-phase data of 156P.}. 
    \label{fig:156_dust}
\end{figure}
Light scattering simulations are executed using the multi-sphere T-matrix code (\textsc{Mstm-V3.0}) and Discrete Dipole Approximations (\textsc{Ddscat-V7.3.3}) for HA and Solids respectively, using the \textsc{Vikram-100 HPC} supercomputing facility at PRL, Ahmedabad. The size range of dust particles used in the model is 1.5$\mu$m - 8.0$\mu$m for HA and 0.12$\mu$m - 3.0$\mu$m for Solids at the wavelength 0.7625$\mu$m which corresponds to the \emph{sloan i} filter.

\section{Discussion}\label{results}
This section will discuss in detail the significance of the results of the spectroscopic, imaging and polarimetric observations of the comet 156P and also consider the results obtained from the modelling of the observed degree of polarization in order to have a deeper understanding of the dust present in the comet.

\subsection{Gaseous emissions}
During the pre and post-perihelion optical spectroscopic observation of the comet, there were clear detection of the emissions from $CN (\Delta \nu = 0)(\lambda3880\AA)$, $C_3 (\lambda4050$\AA), $C_2 (\Delta \nu = +1)(\lambda4690\AA)$ and $C_2 (\Delta \nu = 0)(\lambda5165\AA)$, as illustrated in Fig. \ref{spectra}. The computed production rates at both epochs are as mentioned in table \ref{result_table}. Considering that the perihelion date of the comet was on 2020 November 17, it is observed that the production rates have increased significantly when observed about 5 weeks post-perihelion. Similar increase in production rates, a few weeks after perihelion, has been observed in comet 67P \citep{opitom_67P}. Due to the lack of observational data, it is difficult to remark on the kind of evolution in the production rate which would have occurred in 156P. Still, increase in production rates as observed in the span of 2 months would require an increased level of activity as the comet approached perihelion. Short-period comets tend to have an upper-shield relatively exhausted in the cometary volatile as a result of its multiple passages close to the Sun \citep{comets_in_UV}. Such dust crusts are formed as the comet recedes from the Sun and sublimation ceases. Once the comet returns to the inner Solar system, these crusts can get depleted exposing the layer beneath (fresh volatile material or layers of ice that have different vaporization rates)  \cite{kumar}. 
This could result in an increased activity in the comet in turn giving rise to an increase in the observed production rates of various molecules. But, as pointed out by \citet{marshall_prodrate}, the reason for a significant increase in cometary outgassing cannot be clearly understood without prior knowledge of basic parameters (shape, spin
axis orientation, activity locations) of the comet nucleus.\\
According to \citet{Ahearn_85}, comets with  production rate ratio Q(C$_2$)/Q(CN) $<$ 0.66 are classified as carbon-chain depleted comets and the others are classified as typical comets. The values of the production rate ratio at both epochs given in table \ref{result_table} implies that the comet 156P belongs to the typical class of comets. Also, even after considering the strict definition of carbon depletion in comets as mentioned in \citet{cochran_30years}, the comet can be clearly seen to be belonging to the typical class of comets. Even though a significant difference in the activity of the comet is observed between both the epochs, the production rate ratios are estimated to be consistent. In the same line, the dust-gas ratio (see table \ref{result_table}) is also measured to be consistent during both epochs. This could mean that the comet had possibly formed in a part of the proto-Solar nebula where the availability of volatile components and other building blocks of the comet nucleus were uniform.

\subsection{Dust emission}\label{afrho_discussion}

The amount of dust in the coma is estimated with the help of a proxy parameter, $Af\rho$, as described in Section \ref{dust}. $Af\rho$, computed by the equation.\ref{afrho} is expected to be independent of the aperture size ($\rho$) in the case of a steady state outflow in the coma \citep{Ahearn_Bowel_slope}. In such a case, the profile of the parameter would attain a constant value beyond a particular aperture. Hence, the value of  $Af\rho$ at a larger aperture can be directly used to study the activity of dust production in the cometary coma.  \\
Fig. \ref{afrho_hct} and \ref{afrho_lisa} illustrates the observed profile of $Af\rho$ in the comet, pre and post-perihelion, corresponding to various filter bands. It is seen that on both epochs the profile does not behave as expected in the case of a simple radial outflow model. Hence, it can be inferred that the coma of 156P has a non-steady state dust emission. Also, a possibility of dust grain destruction with nucleocentric distance cannot be ruled out. This results in the fall of column density (N($\rho)$) and hence the $Af\rho$ at larger apertures. Similar kind of behaviour in the dust emission has been observed in a few other short period comets, 221P/LINEAR \citep{afrho_5SPC_5LPC}, 52P/Slaughter-Burnham and 78P/Gehrels  \citep{dustcoma_SPC_afrho_profile} as well as a few long-period comets C/2001 Q4 \citep{C2001Q4_afrho_profile}, C/2000 WM1 \citep{C2000WM1_dust}, C/2010 FB87 and C/2011 L4 \citep{afrho_5SPC_5LPC}. Hence, this behaviour can be considered to be irrespective of the dynamical age of the comet but as a characteristics of the activity in the comet. \\
According to Fig. \ref{afrho_hct}, in the pre-perihelion epoch, the steep rise in the $Af\rho$ profile in the inner coma around 1000 Km might be due to the presence of a compact dust coma \citep[as mentioned in ][]{156P_Atel_Oct}. The gradual decrease in the profile with the increase in nucleocentric distance can either be due to the destruction of a fraction of the dust present in the coma as they move outward or due to a difference in population of dust grains, with different physical properties, between the inner and outer coma \citep[eg., ][]{C2000WM1_dust}. Even though a similar trend in $Af\rho$ profile is observed post-perihelion (see Fig. \ref{afrho_lisa}), it is noticed that the peak which was observed around 1000 km (pre-perihelion),  is observed at around 2000 km (post-perihelion) and has broadened by a few thousand kilometres. This shifting and broadening can be due to an expanded outer boundary of the compact dense inner coma caused by a slightly increased outflow of the dust in the coma owing to the increased activity in the comet. \\
Fig. \ref{afrho_alldate} compares the characteristics of the Af$\rho$ profile over a few selected dates of observation. It is seen that the profile observed on 2020 November 18 is similar to what has been observed on 2020 October 18. As the comet moves in its orbit, away from the Sun, the observed profile is seen to be widening as if the difference in the presence of amount of dust in the  inner and outer coma has reduced. The Af$\rho$ profile observed on 2020 December 8 has a trend is similar to what is observed on 2020 December 15 (see Fig. \ref{afrho_lisa}). Later, it is observed that the peak continues to widen and on December 29 the observed profile is similar to an expected Af$\rho$ profile for a comet with steady state outflow of dust emission. Hence, over the period of observation during pre and post-perihelion epochs, the dust emission in the comet has changed from a strong non-steady state outflow to a steady state. This could be possible if there were certain strong activity in the comet around perihelion which decreased  by the end of December, few weeks after perihelion, as the comet started moving away from the Sun. Detailed dust modelling studies are required to understand the dust distribution in the coma to explain such abnormal Af$\rho$ profiles.

\subsubsection{Coma morphology}\label{morphology}

Dust structures present in the coma can be identified by analysing the coma morphology in broadband filters like R or I since the light observed in these bands directly corresponds to the light scattered by the dust present in the coma. Open/clear filter can be used to get more SNR in case of a faint comet or small telescope, with the only issue being the contamination from the molecular emissions. Still, this technique can be employed to get a general understanding of the coma morphology present in the comet. There are various techniques \citep[eg., ][]{larson_sekanina,samarasinha_process,schwarz_imageprocess,larson_slaughter_process,schleicher_farnham_process} that can be employed to carry out the analysis. In the current work the Larson-Sekanina (LS) processing method proposed by \citet{larson_sekanina} is being used. We use a similar technique followed in \citet{afrho_5SPC_5LPC} for comet 221P, where two images rotated with same angle in the opposite direction are individually subtracted from the original image and then co-added to increase the contrast between any faint structure present in the coma. The angle of rotation is chosen in such a way that the feature is best highlighted. \\
All the images available, as mentioned in table \ref{observations}, few of which are shown in Fig. \ref{original}, were processed using this technique. The selected outputs of the LS processing are illustrated in Fig. \ref{ls_processed} with the angle of rotation used for the processing and directions of Sun and heliocentric velocity marked in all the panels. In panel (a), which is the only observation before perihelion, the presence of a strong feature, marked J1, probably a part of the main dust tail, is clearly visible along with a few faint features. As the comet moved in its orbit, there is a projection effect which causes a flip in the position of J1. Two other strong features, J2 and J3, are detected in the rest of the images indicating an increase in activity of the comet around perihelion. It is also observed that the jets, J2 and J3, remains active in December (see panel (c) for observation on 2020 December 08). This increase in activity could explain the increase in the production rates when the comet was observed on 2020 December 15.\\
Even if it is not demonstrable here, the prominent feature J1 can be suspected to be a part of the dust tail arising from an active region facing the Sun, being bent into the tail direction by the Solar radiation pressure. \citet{farnham_JFCmorphology} named such jets which appear unchanged for a long period of time as fixed jets. These fixed jets are usually centred on the rotation axis. According to the collimated jet model \citep{sekanina_collimatedjetmodel}, isolated active regions producing such fixed jets can produce spatial variations in the density or particle size of the dust particles present in the coma. Detailed measurement of the comet's position along its orbit, analysis of dust particle size etc are required to confirm this assumption.  \\
Comparing the panel (b) with the other panels in Fig. \ref{ls_processed}, it can be observed that the jets are more compact and streamlined during the initial epochs of observation =with the angle between \textit{J1 and J2} being only 25.7$^\circ$. In the later epochs the coma opens up or the jets move apart systematically with the angles between \textit{J1 and J2} being 49.9$^\circ$ and 59.7$^\circ$ on 2020 December 08 (Fig.\ref{ls_processed} panel (c)) and 2020 December 20 (Fig.\ref{ls_processed} panel (d)) respectively. This angular variation between the observed jets more or less corresponds to the epochs when the Af$\rho$ profiles are observed to start widening (see Af$\rho$ profile corresponding to 2020-12-08 illustrated in Fig.\ref{afrho_alldate}). Hence, there is a possibility that, such a process would have resulted in an increased outflow of dust into the outer coma causing the widening of Af$\rho$ profile peak as seen in Fig. \ref{afrho_lisa}. From Fig. \ref{ls_processed}, it can be observed that the jets which were most prominent during the first half of December starts to loose strength and subsides towards the end of December. Fig. \ref{afrho_alldate} illustrates that the Af$\rho$ profile started widening as the coma/jets opened up during the beginning of December and then attained a normal profile
%similar to what is usually expected in a comet having a steady state outflow of dust 
towards the end of December, once the jets subsided significantly. Hence, in the current case it can be inferred that the non-steady state outflow of dust resulting in the abnormal Af$\rho$ profile could be a result of the presence of strong dust jets arising from the nucleus. As the prominence of the jets subsided, the dust outflow would have tended more or less towards a steady state, reducing the spatial variation in the coma, causing the Af$\rho$ profile to attain a constant value independent of the aperture size, as it is normally expected.

\subsection{Polarization and dust properties}

\begin{figure}[h]
	\centering
	\includegraphics[width=0.7\linewidth]{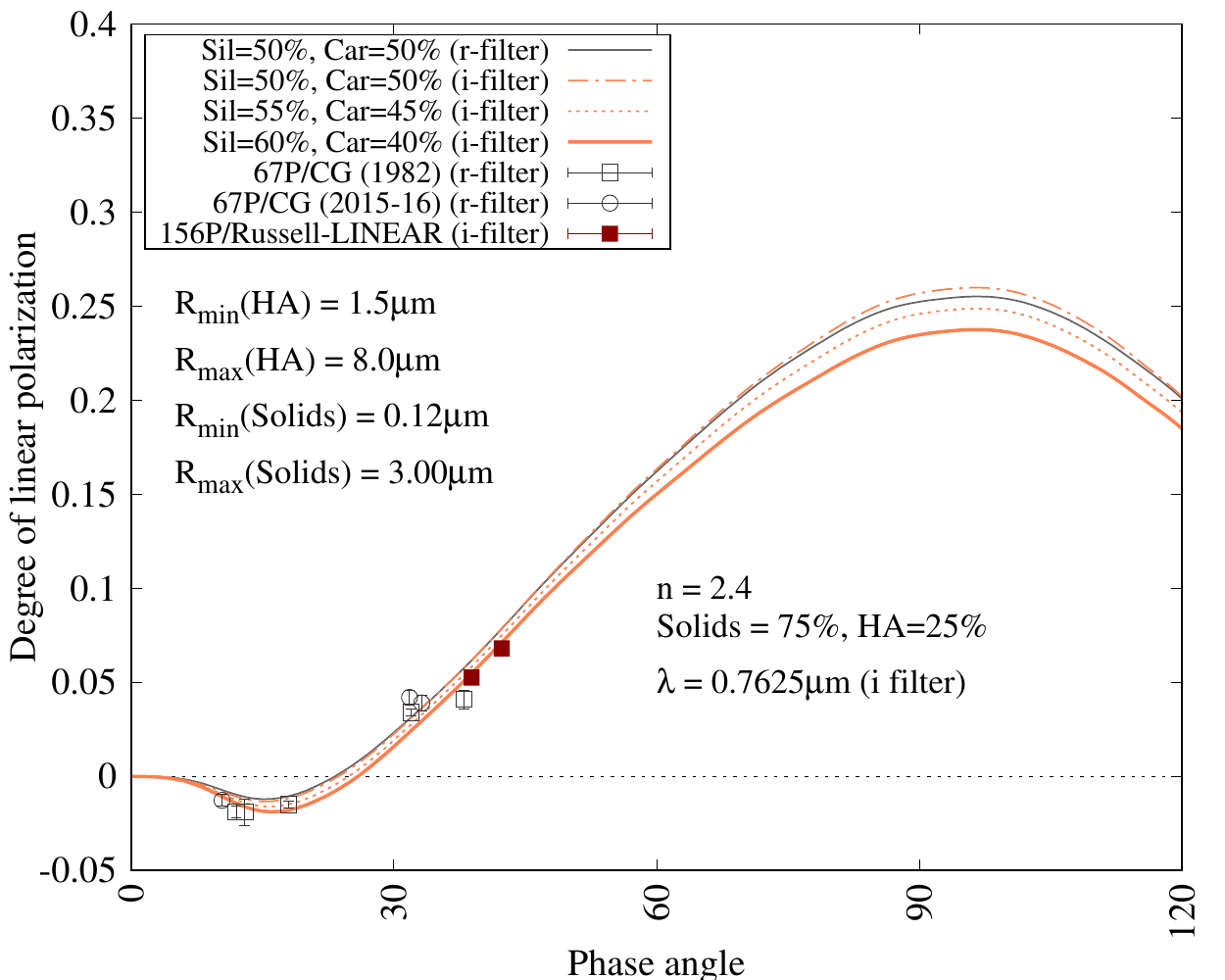}
    \caption{ \label{fig:156_fit} Best fit results for the comet 156P/Russell-LINEAR with the observations (dark-red solid squares). The light-red coloured lines indicate the best fit modelled polarization-phase curves obtained from this study in the \emph{sloan i} 
    filter and the grey solid line indicate the best fit model obtained for 67P from \citet{pritish_modelling}. The grey hollow squares/circles denote the observations of 67P in $R$ filter \citep{67pc-g1, 67P_pol}.} 
   
\end{figure}
\begin{figure}[h]
  \centering
  \includegraphics[width=0.9\linewidth]{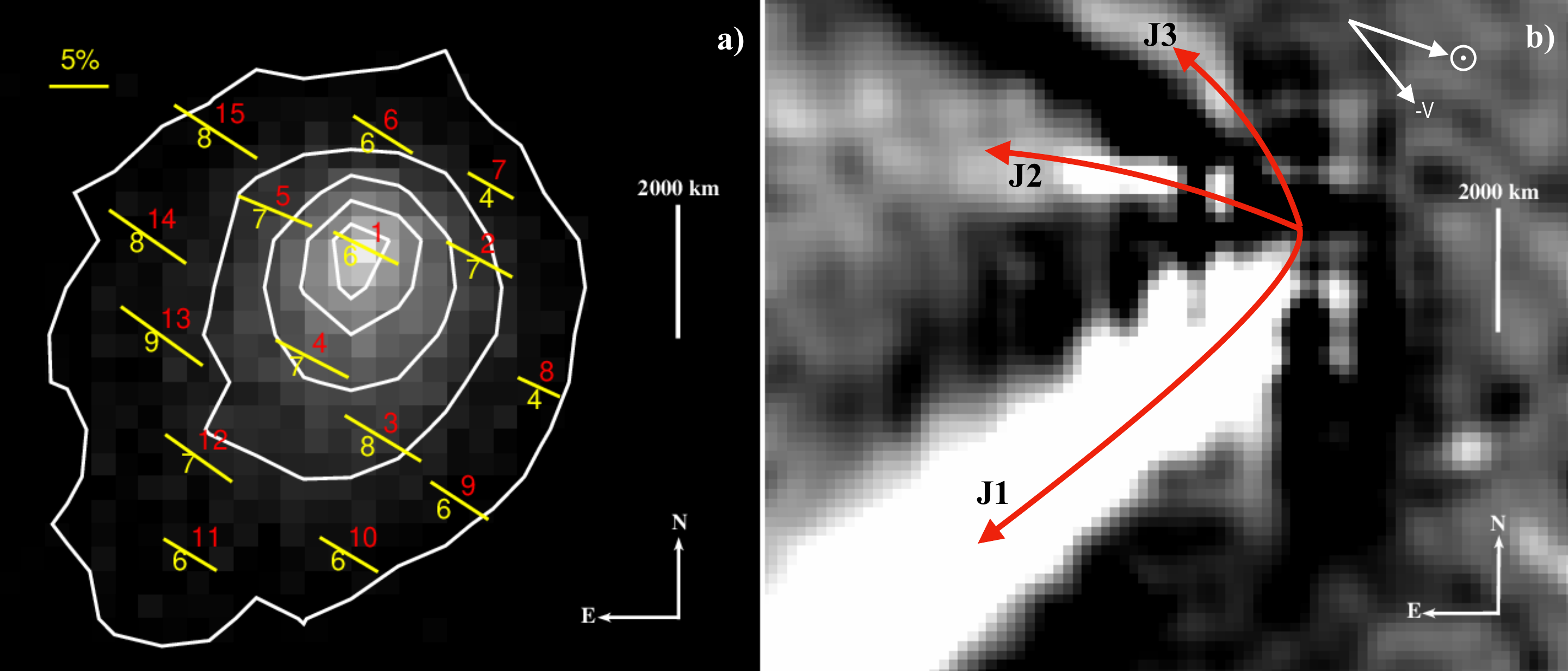}
  \caption{a) Illustration of spread of polarization vectors in the coma of the comet 156P observed on 2020-12-16. The direction of the local polarization plane is indicated by the orientation of the vectors, and their length denotes the degree of polarization. Numbers in yellow represent the degree of polarization rounded off to integer and red represents the aperture number (see table \ref{location_pol}). The contours simply represent the intensity variation in the coma. The vertical solid white line depicts the physical scale on the sky at the comet. b) Illustration of the observed jet directions in the LS processed image obtained on 2020-12-18.}
              \label{pol_loc}
    \end{figure}

%\cite{pritish_modelling}
The degree of polarization and polarization angle were measured for the comet at two different phase angles. The observed values are tabulated in table \ref{comet_pol_result}. Dust modelling as described in section \ref{model} is an effective method to understand the possible composition and characteristics of the dust particles present in the comet. Fig. \ref{fig:156_fit} shows the best fit modelled data for the observed polarimetric response obtained for comet 156P at the \emph{sloan i} filter for 75 percent Solids + 25 percent HA having power-law size distribution $n$ = 2.4. The modelled data indicates the presence of high amount of silicate/low absorbing material (50  to 60 percent) as obtained in the case of 67P \citep{pritish_modelling}. 
The polarimetric study of 156P reveals dust properties similar to those obtained in the case of the comet 67P. The shallow polarimetric slope indicates the presence of highly processed large size dust particles. The short orbital period allows frequent weathering of the dust reducing the amount of small fractals leaving behind the solids with low porosity in the coma. \\
It was observed that there are strong fixed jets present in the coma. As mentioned earlier, presence of such jets can create spatial variation in the particle size of the dust, which can produce a spatial variation of the degree of polarization in the coma. Since, the presence of any systematic change in degree of polarization with increasing aperture was already ruled out, an analysis was carried out on the 2020 December 16 polarimetric data in order to explore the possibility of difference in degree of polarization at different locations in the coma due to the presence of the  strong jets. The polarimetric data obtained for 2020 November 13 could only be used for computing  the total coma polarization as the comet was not bright enough to determine the polarization at different locations within the coma. As shown in Fig. \ref{pol_loc} panel (a), for the 2020 December 16 data, various locations in the coma were chosen to compute the corresponding polarization for a circular aperture of 2 pixel (706 km) diameter. Similar computational techniques as mentioned in Section \ref{pol_anal} were employed at all the locations. The degree of polarization corresponding to each location is illustrated in the form of polarization vectors as well as rounded off to integers and mentioned at each location in yellow. The actual observed degree of polarization is also  mentioned in table \ref{location_pol}. Even though the errors are larger as compared to the earlier analysis, it is noticed that there is a variation in the observed degree of polarization from point to point. At the same time, the jet features can not be distinguished in the polarimetric image due to the low resolution of the polarimeter . Hence, the jet features can be expected to be similar to what is observed on 2020 December 18 (see panel (b) of Fig. \ref{pol_loc}).\\
A pattern in the variation of polarization can not be definitively concluded due to the large errors in the observed values. It is observed that the degree of polarization computed at the points 3, 4, 5, 12, 13, 14 and 15 are distinctly higher than what is observed at the other points. Upon careful comparison with panel (b) in Fig. \ref{pol_loc}, it is spotted that these points lie in the area dominated by the observed strong jets \textit{J1, J2, J3}. The presence of such jets would have created a localised difference in the physical characteristics of the dust population giving rise to a difference in the observed polarization \cite[eg., ][]{67P_pol}. The degree of polarization is observed to be lower a few thousand kilometres away from the nucleus.
Even though it is not a conclusive evidence, it can be indirectly presumed that, despite the jets being strong, their dominance is only in the inner coma within a few thousand kilometres. The dust particles arising from these jets could be getting destroyed as they move into the outer coma (giving rise to the dip in the Af$\rho$ profile as discussed in Section \ref{afrho_discussion}) causing the observed polarization to be lower.

\begin{table}[h]
    \centering
    \caption{Degree of polarization corresponding to the locations displayed in Fig. \ref{pol_loc} panel (a).}
    \begin{tabular}{|c|c|c|}
\hline
  \multicolumn{1}{|c|}{Aperture number} &
  \multicolumn{1}{c|}{$p_{obs}$ ($\%$)} & \multicolumn{1}{c|}{$\theta_{corr}$ ($^\circ$)}\\%\textit{p$_{obs}$}} \\
\hline
  1 & 6.11 $\pm$ 0.41 & 154 $\pm$ 4\\
  2 & 6.63 $\pm$ 0.46 & 152 $\pm$ 4\\
  3 & 7.52 $\pm$ 0.45 & 149 $\pm$ 4\\
  4 & 6.91 $\pm$ 0.49 & 153 $\pm$ 4\\
  5 & 6.74 $\pm$ 0.47 & 158 $\pm$ 5\\
  6 & 5.95 $\pm$ 0.50 & 148 $\pm$ 5\\
  7 & 4.44 $\pm$ 0.52 & 151 $\pm$ 4\\
  8 & 3.91 $\pm$ 0.56 & 155 $\pm$ 5\\
  9 & 5.87 $\pm$ 0.58 & 147 $\pm$ 5\\
  10 & 5.71 $\pm$ 0.62 & 149 $\pm$ 5\\
  11 & 5.52 $\pm$ 0.62 & 149 $\pm$ 5\\
  12 & 6.90 $\pm$ 0.60 & 145 $\pm$ 5\\
  13 & 8.52 $\pm$ 0.50 & 144 $\pm$ 5\\
  14 & 7.93 $\pm$ 0.55 & 145 $\pm$ 5\\
  15 & 8.34 $\pm$ 0.48 & 147 $\pm$ 5\\
\hline\end{tabular}
    
    \label{location_pol}
\end{table}

\section{Conclusions}
In this work, we have studied the short period Jupiter family comet 156P/Russell-LINEAR using spectroscopic, photometric and polarimetric techniques with the help of HCT, MIRO and IAC-80 telescopes. Emissions from $CN (\Delta \nu = 0)$, $C_3 (\lambda4050$\AA), $C_2 (\Delta \nu = +1)$ and $C_2 (\Delta \nu = 0)$ are detected on both epochs of spectroscopic observation. The observational results from spectroscopy, imaging and polarimetry along with dust modelling help us in arriving at the following conclusions:

\begin{enumerate}
    \item The production rates observed for CN, C$_2$ and C$_3$ are comparable to those observed for Jupiter family comets. From the spectroscopic results obtained for the two epochs of observation, an increase in activity is observed post perihelion.
    \item The value of Q(C$_2$)/Q(CN) evidently classifies the comet as a typical comet.
    \item  The images processed through Larson-Sekanina technique, indicating the presence of strong jets throughout the observational period. This, along with the observed increase in production rates, suggests that there has been an increase in the activity as the comet crossed perihelion.
    \item A non-steady state outflow of dust emission is observed during the initial epochs of observation. The jets which are arising from the isolated active regions in the comet nucleus can create spatial variation in the density or particle size of the dust present in the coma generating an abnormality in the observed Af$\rho$ profiles.
    \item Towards the end of 2020 December, about 6 weeks after perihelion, the dust jets are observed to subside, which could have reduced the spatial variations in the coma. This may be a possible reason for the Af$\rho$ profiles to tends towards a general profile as in the case of a steady state outflow of dust.
    \item The degree of polarization of the comet obtained on both the epochs are in good agreement with the values observed for Jupiter family comets at similar phase angles. 
    \item A variation in the degree of polarization at various locations in the coma, dominated by the dust jets,  has been observed. This suggests a probability of dust jets being a prominent reason for producing localised differences in the density or size of the dust particles present in the coma.
    \item The best fit model for the observed polarization of the comet 156P suggests the presence of higher percentage ($>$ 50$\%$) of silicates or low absorbing material. The model also indicates the coma to be dominated by higher amount of larger particles with lower porosity, having a power law size distribution index $n$ = 2.4. This can be due to the short orbital period and/or frequent weathering of the smaller materials with high porosity similar to what is seen in the case of the comet 67P. 
    
    \end{enumerate}

  The comet 156P/Russell-LINEAR has not been well studied in any of its previous apparitions due to its unfavourable orbital positioning. This restricts us from generalising the characteristics of the comet with our set of data or to compare it with other data sets. However, the behaviour of the activity in the comet around perihelion and the modelled dust properties obtained with the help of polarization data, points to a similarity to another Jupiter family comet 67P. Further study of the comet in its upcoming favourable apparitions would be welcome to have a better understanding of this comet.

\section*{Acknowledgements}

We acknowledge the local staff at the Mount Abu Observatory for their help. We thank the staff of Indian Astronomical Observatory, Hanle and Centre For Research \& Education in Science \& Technology, Hoskote that made these observations possible. The facilities at IAO and CREST are operated by the Indian Institute of Astrophysics, Bangalore. \\Work at Physical Research Laboratory is supported by the Department of Space, Govt. of India. The authors
acknowledge the use of the supercomputing facility Vikram-HPC
at PRL, Ahmedabad, where all the parallel computations related to the dust modelling in this work were executed.\\
This article is partially based on observations made in the Observatorios de Canarias del IAC with the IAC-80 telescope operated on the island of Tenerife by the Instituto de Astrofísica de Canarias in the Observatorio del Teide.\\
We also thank the unknown referees whose valuable comments have significantly improved the quality and readability of the paper.
This research made use of Astropy\footnote{http://www.astropy.org}, a community-developed core Python package for Astronomy \citep{astropy:2013, astropy:2018}.

 \bibliographystyle{elsarticle-harv} 
 \bibliography{156P_icarus.bib}

\appendix
\section{Solar continuum removal}
\label{continuum_removal}

\begin{figure}[h]
\centering
\subcaptionbox{Comet spectrum overplotted with the scaled spectrum of the solar analog star. \label{comet_solar}}%
  [0.49\linewidth]{\includegraphics[width = 1\linewidth]{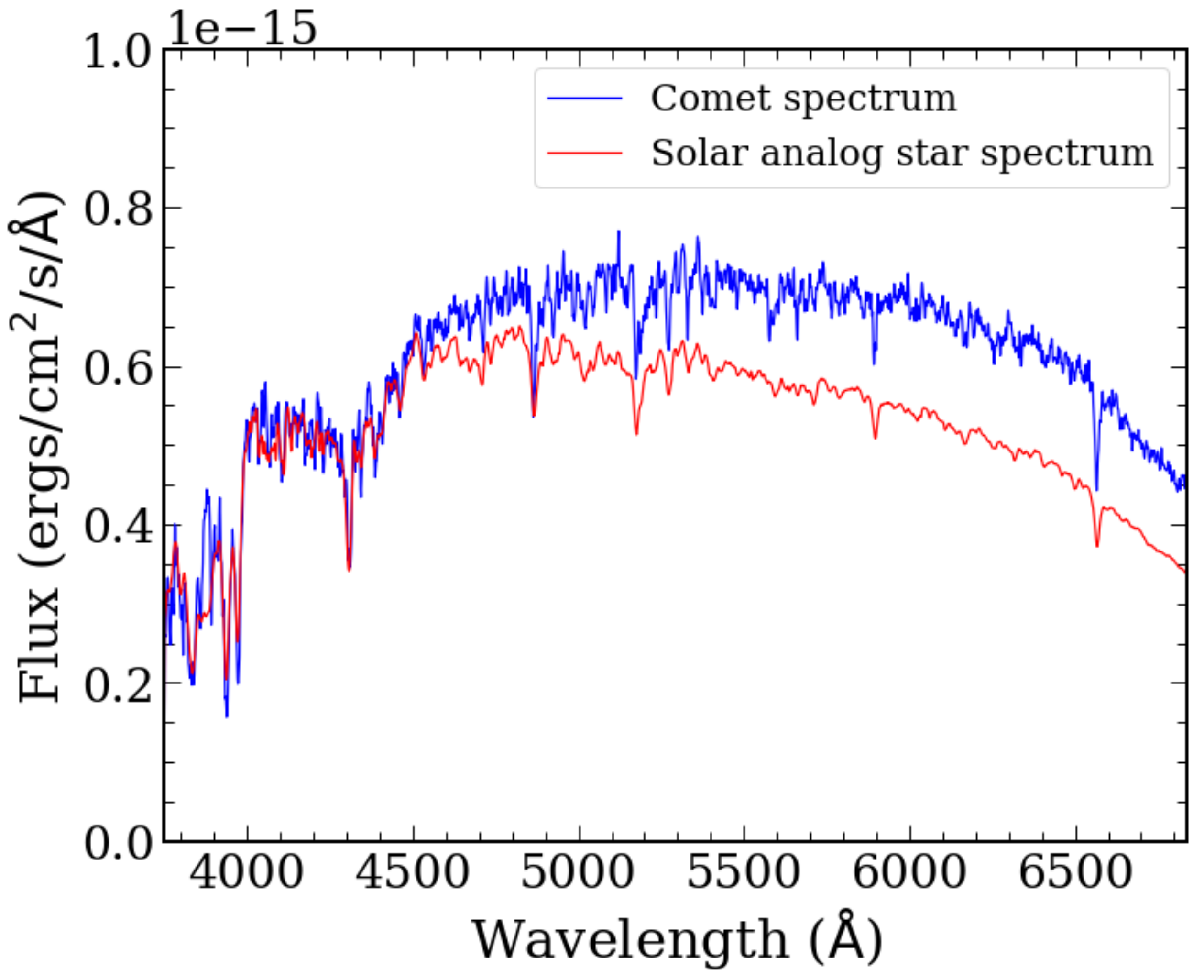}}
  \hfill
\subcaptionbox{Comet spectrum overplotted with the synthetic continuum spectrum. \label{continuum_fitted}}
  [0.49\linewidth]{\includegraphics[width = 1\linewidth]{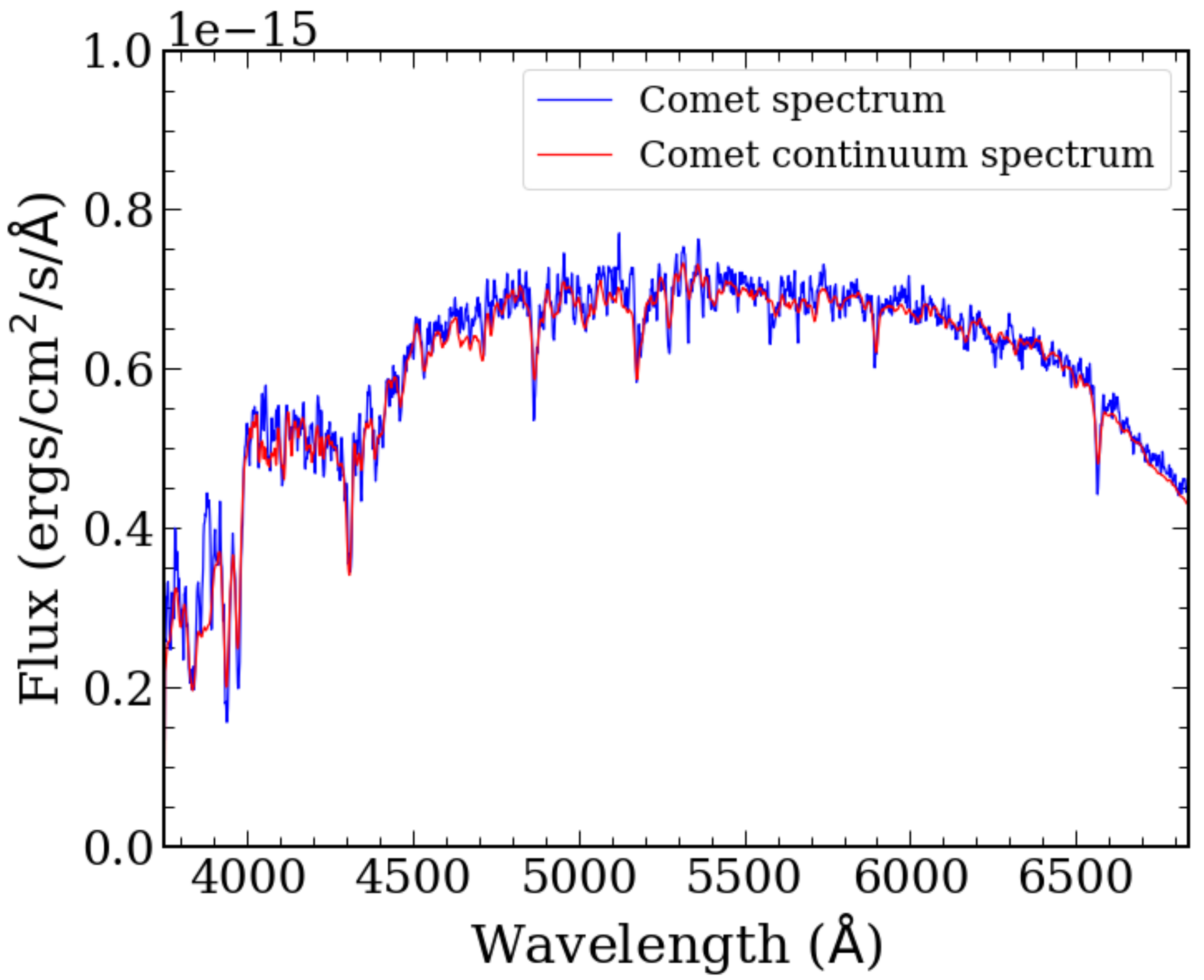}}
\caption{Continuum removal of comet spectrum extracted at the photocentre using a single pixel (103 km) aperture.}
\end{figure}

 The comet spectrum is comprised of both the coma gas emission spectrum and the spectrum of the sunlight scattered by the dust particles. Hence, it is necessary to remove the continuum signal in order to study the gaseous emissions. A Sun spectrum or a solar analog spectrum is used to remove the continuum signal. Here in this work, a solar analog star has been used. Initially, the observed solar analog spectrum is normalised and scaled to the comet continuum flux (see Fig. \ref{comet_solar}). A polynomial is fit to both the comet and solar analog spectrum for the continuum windows mentioned in \citet{C2019Y4_continuumremoval}. The scaled solar analog star spectrum is multiplied by the ratio of these polynomials (to correct for the redder nature of the cometary dust) in order to obtain a continuum spectrum of the comet as shown in Fig. \ref{continuum_fitted}. This continuum spectrum is now subtracted from the original comet spectrum to obtain the pure emission spectrum.

\end{document}